%% file: main.tex
\documentclass[12pt,oneside]{book}

\usepackage[english]{babel}                 
\usepackage{Custom}                         
\usepackage{AuxiliaryFiles/AuxiliaryFiles}  
\usepackage{Custom-2}                       

\usepackage[natbib,style=authoryear,backend=biber, giveninits=true]{biblatex}
\DeclareNameAlias{sortname}{family-given}

\begin{document}
\input{Structure/Title.tex} 

\newcommand\IndexYes{1}
\input{AuxiliaryFiles/Frontmatter}  
\input{Structure/Introduction}      

\input{AuxiliaryFiles/Mainmatter}

\input{Structure/Session1}
\input{Structure/Session2}
\input{Structure/Session3}
\input{Structure/Session4}
\input{Structure/Session5}
\input{Structure/Session6}
\input{Structure/Session7}
\input{Structure/Session8}
\input{Structure/Session9}
\input{Structure/Session10}

\input{Structure/Session11}
\input{Structure/Session12}
\input{Structure/Session13}
\input{Structure/Session14}
\input{Structure/Session15}
\input{Structure/Session16}

\end{document}

%% file: Structure/Title.tex
\begin{titlepage}
\begin{figure}[ht]\centering
        \includegraphics[width=.8\textwidth]{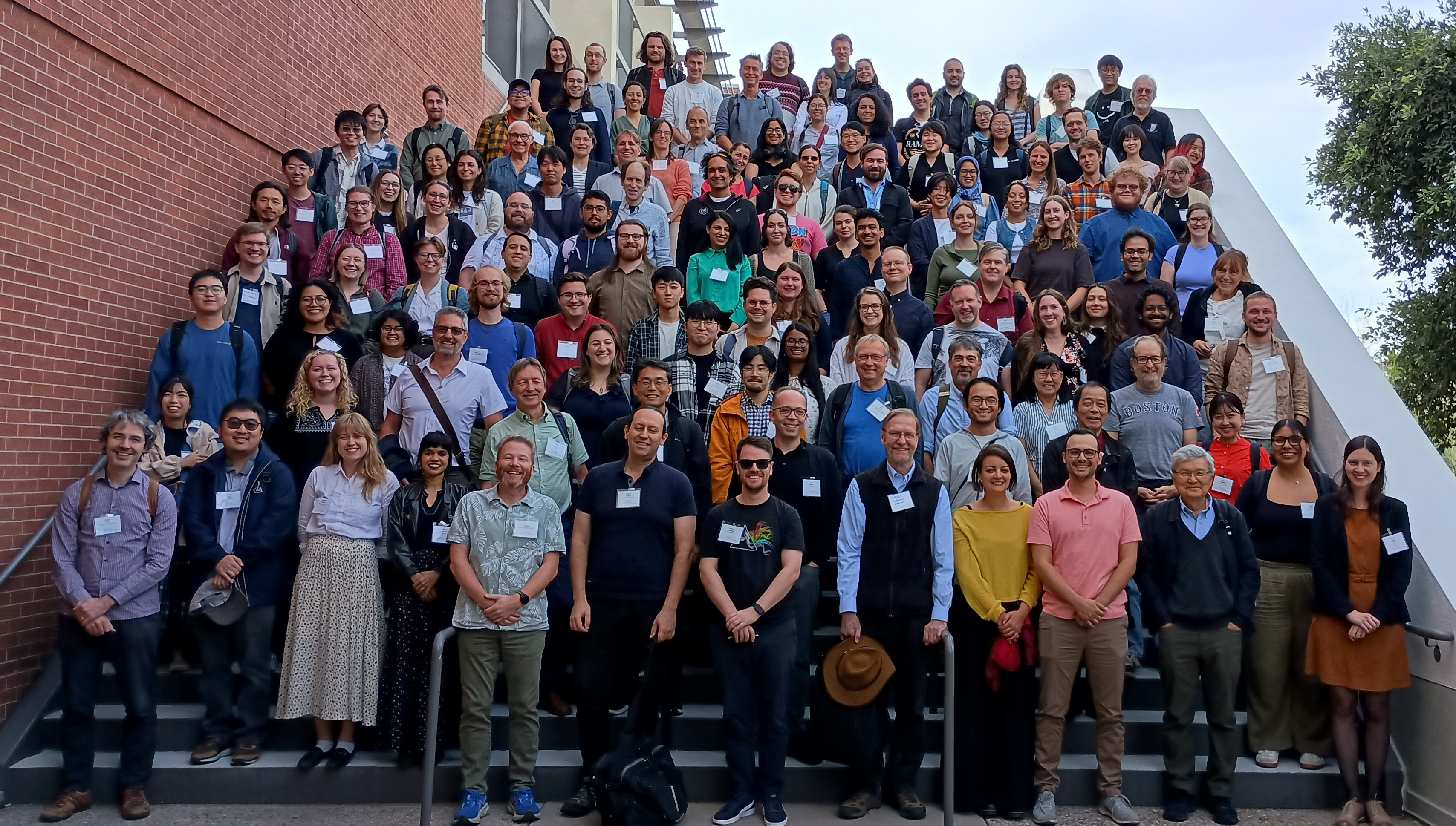}
\end{figure}

\begin{center}
\parbox[b]{\textwidth}{
    \centering
	{\HUGE\bfseries The Dusty Universe 2025}                          \\
	{\Large {\textbf{\textit{The Fifth Pandust Conference}}\\[5pt] \large Tucson, Arizona, USA -- November 10-14, 2025}}}
    \end{center}
\vspace{0.4cm}

\begin{minipage}[t]{.5\textwidth}
{\Large\bfseries {Discussion moderators}}              \\[10pt]
        {\large
       \textsc{Mikako Matsuura}\\[4pt]
        \textsc{Elizabeth Tarantino}\\[4pt]
        \textsc{Meriem Elyajouri}\\[4pt]
        \textsc{Irene Shivaei}\\[4pt]
        \textsc{Thomas Henning}\\[4pt]
        \textsc{Maarten Baes}\\[4pt]
    \textsc{Andrew Saydjari}\\[4pt]
    \textsc{Chian-Chou Chen}\\[4pt]
        \textsc{Aki Takigawa}\\[4pt]
    \textsc{Jacob Jencson}\\[4pt]
    \textsc{Thomas Lai}\\[4pt]
    \textsc{Frédéric Galliano}\\[4pt]
    \textsc{Samir Salim}\\[4pt]
    \textsc{Jed McKinney}\\[4pt]
    \textsc{J.D. Smith}\\[4pt]
         }
\end{minipage}
\begin{minipage}[t]{.5\textwidth}\raggedleft
{\Large\bfseries {Discussion organizers}}              \\[5pt]
        {\large
        \textsc{Marjorie Decleir}              \\
        \textsc{Annalisa De Cia}               \\
        \textsc{Brandon Hensley}               \\

}

\end{minipage}

\vfill
\begin{center}
\large\today
\end{center}
\end{titlepage}

\let\cleardoublepage\clearpage 

%% file: AuxiliaryFiles/Frontmatter.tex
\frontmatter

\newgeometry{
	top=3cm,
	bottom=2.5cm,
	left=3cm,
	right=3cm,
	marginparsep=0cm,
	marginparwidth=0cm,
	voffset=0pt,
	hoffset=0pt,
	headheight=0pt,
	headsep=0pt,
	footskip=14pt,
	footnotesep=0pt
}

\fancypagestyle{fancyfront}{
	\renewcommand{\headrulewidth}{0pt}
	\renewcommand{\footrulewidth}{0pt}
	\pagenumbering{Roman}
	\fancyhead[L,R]{}
	\fancyfoot[C]{\small\thepage}
	\fancyfoot[LO,RE]{}
	\fancyfoot[LE,RO]{}}

\pagestyle{fancyfront}

\ifnum\IndexYes=1
\contentsmargin{0cm}
\titlecontents{part}[0pc]
{\addvspace{8pt}}{}
{\fontsize{12}{8}\bfseries}
{\tiny\color{black!50}\;\;\dotfill\;\fontsize{12}{8}\bfseries\color{black}\PageName\, \thecontentspage}
\titlecontents{chapter}[0pc]
{\addvspace{16pt}\fontsize{12}{8}\bfseries\thecontentslabel}{\hspace{.3cm}}{}
{\tiny\color{black!50}\;\;\dotfill\;\fontsize{12}{8}\bfseries\color{black}\PageName\, \thecontentspage}
\titlecontents{section}[2pc]
{\addvspace{2pt}\bfseries\thecontentslabel\normalfont}{\hspace{.5cm}}{}
{\tiny\color{black!50}\;\;\dotfill\;\normalsize\color{black}\PageName\, \thecontentspage}
\titlecontents{subsection}[6pc]
{\addvspace{2pt}\bfseries\thecontentslabel\normalfont}{\hspace{.5cm}}{}
{\tiny\color{black!50}\;\;\dotfill\;\normalsize\color{black}\PageName\, \thecontentspage}

\begin{flushright}
        \HUGE\bfseries\ContentsName
\end{flushright}

{
    \let\cleardoublepage\relax
	\renewcommand\contentsname{}
    \vspace{-4cm}
	\tableofcontents
	\vspace{.25cm}
}
\fi

\restoregeometry

%% file: Structure/Introduction.tex

\chapter*{}\addcontentsline{toc}{part}{\IntroductionName}
\vspace{-5cm}
\begin{flushright}
    \HUGE\bfseries\IntroductionName
\end{flushright}

\noindent ``The Dusty Universe: The Fifth Pandust Conference'' took place in Tucson (AZ, USA) from November 10 until 14, 2025. This was the fifth meeting in the series of dust meetings that happen (approximately) every 5 years. The previous meetings were called ``Astrophysics of Dust'' (Estes Park, Colorado, USA, 2003), ``Cosmic Dust - Near and Far'' (Heidelberg, Germany, 2008), ``The Life Cycle of Dust in the Universe'' (Taipei, Taiwan, 2013), ``Cosmic Dust: Origin, Applications \& Implications'' (Copenhagen, Denmark, 2018).

The goal of this meeting was to get the dust community together to review where we are, hear exciting new results, and make plans for the future. The meeting encompassed all aspects of dust investigations including observations, theory, modeling, and laboratory studies. The conference consisted of invited review talks, contributed talks and posters. Science topics included interstellar dust (Milky Way \& nearby galaxies), circumstellar dust (including Solar System \& exoplanets), dust in Galaxies (including high-z), lifecycle of dust, and future needs (laboratory, theory, \& observations), with a particular focus on results from JWST and ALMA, and on nanodust (including PAHs). The meeting was set up to be accessible to all levels including graduate students. The full program, a selection of posters and presentation slides, and the list of participants can be found on the meeting website: \url{https://pandust2025.sciencesconf.org}.

The organization of this meeting was chaired by Karl Gordon (STScI, Baltimore, USA), and the Scientific Organizing Committee (SOC) consisted of a mix of senior and junior astronomers from the dust community from different institutes and universities across the USA, Europe and Australia. The SOC members were 
Anja Andersen (University of Copenhagen, Denmark), Alessandra Candian (University of Amsterdam, Netherlands), Annalisa De Cia (ESO, Garching, Germany), Marjorie Decleir (ESA/STScI, Baltimore, USA), Frédéric Galliano (CEA Paris-Saclay, France), Thomas Henning (MPIA, Heidelberg, Germany), Brandon Hensley (JPL/Caltech, USA),
Francisca Kemper (ICE-CSIC, Barcelona, Spain), Karl Misselt (University of Arizona, USA), Darach Watson (University of Copenhagen, Denmark), Adolf Witt (University of Toledo, USA), and Tayyaba Zafar (Macquarie University, Australia). 

On November 12\textsuperscript{th}, we organized breakout discussion sessions covering a wide range of interesting dust-related topics. We had two sessions of about one hour each, with 8 parallel discussions in the first, and 7 in the second session. Each discussion was led by a moderator, and at the end, each moderator presented a short oral summary of their discussion session to all conference participants.

The purpose of this document is to capture the main topics/questions that were discussed, the key conclusions of these discussions, the challenges and possible solutions that were brought up, and the open questions that still remain to be answered. We hope that this document records our findings and challenges for the future generation (for example the next dust meeting in 5 years).\\

%% file: AuxiliaryFiles/Mainmatter.tex
\mainmatter

\titleformat{\chapter}[display]{\bfseries\Large}	{\filleft\MakeUppercase{Discussion session} \HUGE\thechapter}{.4ex}{\vspace{1ex}\filleft\HUGE}[\vspace{1ex}]
\titlespacing*{\chapter}{0pt}{0.1\baselineskip}{0.5\baselineskip}

%% file: Structure/Session1.tex
\begin{refsection}
\chapter{Dust formation processes}

\textbf{Moderator: Mikako Matsuura}

\section{Context}
Dust grains are ubiquitous throughout the interstellar medium (ISM) and are detected in galaxies from the Milky Way \citep[e.g.][]{1998ApJ...500..525S, 2011A&A...536A..19P, 2019ApJ...887...93G} to high-redshift systems \citep[e.g.][]{2015Natur.519..327W, 2022MNRAS.515.3126I, witstok23}. However, the processes responsible for the initial formation of dust in astrophysical environments remain uncertain.

\section{Dust formation sites and global dust budgets}
The circumstellar envelopes of Asymptotic Giant Branch (AGB) stars are well-established sites of dust formation \citep[e.g.][]{1996A&ARv...7...97H, 2005pcim.book.....T}. However, recent studies indicate that multiple sources contribute to the dust budget in the ISM. In addition to AGB stars, supernovae \citep{1981ApJ...248..138D, 1999ApJ...521..234A,  2011Sci...333.1258M, 2012ApJ...760...96G} and, more recently, novae have been identified as alternative dust formation sites \citep{1994Natur.369..539S, 2018ApJ...858...78G, 2001ApJ...551.1065A}.
There is growing evidence that grain growth within the ISM may substantially increase the total dust mass \citep[e.g.]{Draine2009}. However, this process requires pre-existing seed grains (i.e., stardust), since grain growth proceeds through the accretion of atoms and molecules onto existing dust particles.

In this discussion, we distinguish between primary dust formation (i.e., condensation of dust from the gas phase) and grain growth as separate processes.

\subsection{What do we know about the populations that produce dust grains?}

The first two questions we address are:
\begin{itemize}
    \item How can we constrain the amount of dust formed by a wide range of dust sources?
    \item How much dust is present in the ISM, and how can it be constrained observationally?
\end{itemize}

Pre-solar grains identified in meteorites are predominantly attributed to AGB stars \citep{2003TrGeo...1...17Z,  2008ApJ...682.1450N}. However, this apparent dominance is likely affected by survival bias, since only grains that withstand processing in the ISM, survive incorporation into the protoplanetary disk, and are ultimately accreted into asteroidal material, are preserved \citep{2025LPICo3090.1649S}.
A minority of pre-solar grains exhibit isotopic and structural signatures consistent with mixed origins, reflecting both stellar condensation and subsequent ISM processing.
By contrast, most material in asteroids is not pre-solar but instead consists of Solar System dust that was thermally processed and recondensed within the early solar nebula.

\subsection{Recent progress in the dust budget census}

The dust budgets of the Large and Small Magellanic Clouds have been quantified \citep{2009MNRAS.396..918M, 2012ApJ...748...40B}, as has that of the Milky Way \citep{Jura1989, Bertre2005}.
However, the Milky Way dust budget remains uncertain due to systematic effects, particularly uncertainties in stellar distances and ISM extinction toward the Galactic plane. The former is expected to improve with ongoing analyses based on Gaia data \citep{Gaiamission2016, 2019ApJ...887...93G, Bailer-Jones2021, 2022MNRAS.512.1091S}.
Recent JWST imaging surveys have further advanced studies of AGB stars in Local Group galaxies \citep{2015ApJS..216...10B, 2025ApJ...991...24B}.

Recently, additional potential dust sources/processes have been identified:
\begin{itemize}
    \item Binarity can significantly impact dust formation (e.g., WR stars, red novae).
    \item Merger events as potential dust formation sites.
\end{itemize}

High pressures can arise from interactions in binary systems, providing the conditions necessary for dust condensation, as observed in some Wolf-Rayet (WR) binaries \citep{2020MNRAS.498.5604H, 2022NatAs...6.1308L}.
The ejecta of red novae expand more slowly than those of supernovae, typically at velocities of a few hundred km\,s$^{-1}$ \citep{2018A&A...617A.129K}. As a result, shock destruction may be less severe, potentially favouring dust survival.
Recurrent novae undergo eruptions on timescales of order a decade. Over long timescales, repeated outbursts may contribute non-negligibly to dust production \citep{2023ApJ...954L..16B}.
Stellar mergers are also potential dust sources. Their ejecta are often oxygen-rich and are therefore expected to condense silicate species such as olivine \citep{2026ApJ...999...16K}.

\subsection{Metallicity effects}

In AGB stars, metallicity plays an important role: at lower metallicities, carbon-rich (C-rich) AGB stars dominate dust production \citep{2009Sci...323..353S, 2009MNRAS.396..918M, 
2013MNRAS.429.2527M,  2025ApJ...991...24B}. In contrast, in supernovae, elements are synthesized in situ and subsequently condense into dust grains \citep{2003ApJ...598..785N, 2015A&A...575A..95S, 2018MNRAS.480.5580S}. As a result, the metallicity of the progenitor gas has little impact on dust formation, at least at currently detectable levels; to date, no clear metallicity dependence in elemental synthesis has been observed.

\subsection{How to form PAHs?}

Both bottom-up and top-down processes contribute to dust and PAH formation \citep{1989ApJS...71..733A, 2018Galax...6..101C}.
These processes are likely not universal across different environments, including molecular clouds, the diffuse ISM, photodissociation regions (PDRs), and AGB envelopes \citep{2025MNRAS.540.1984C}.
PAHs in the diffuse ISM tend to be smaller than those in denser regions. In denser environments, enhanced turbulence may increase grain-grain collisions \citep{Peeters2024}, promoting coagulation rather than shattering. This raises the possibility that some PAHs are formed in situ through coagulation processes \citep{2024MNRAS.528.3392W, 2025MNRAS.542.1287M}.

\subsection{How are complex silicates formed beyond SiO$_2$}

Dust chemical modelling requires non-equilibrium chemistry, and significant uncertainties and challenges remain \citep{1999A&A...341L..47D, 2013ApJ...776..107S}.

\printbibliography[heading=subbibliography]
\end{refsection}

%% file: Structure/Session2.tex
\begin{refsection}
\chapter{Interpreting PAH emission}

\textbf{Moderator: Elizabeth Tarantino}\\

The mid-infrared (MIR) spectrum of interstellar material is dominated by broad emission features attributed to vibrational modes of small carbonaceous dust grains commonly referred to as polycyclic aromatic hydrocarbons (PAHs). This discussion session was motivated by the need to establish a common framework for interpreting PAH emission in the new JWST era as well as the connection between observations, models, and laboratory data.

\section{PAH continuum subtraction in JWST photometric filters}
The infrared spectral coverage, sensitivity and spatial resolution of JWST are transforming our understanding of PAH emission in the local universe and beyond. However, a key limitation of JWST is the field of view (FOV) of its workhorse spectroscopic instruments; the NIRSpec and MIRI integral field units (IFUs) have FOVs of only $3-5^{\prime\prime}$, making it impossible to map nearby galaxies efficiently. Fortunately, PAH emission in the local universe can be observed using three dedicated photometric filters: F335M on NIRCam and F770W and F1130W on MIRI. Interpreting PAH emission measured in these filters requires reliable prescriptions for estimating and subtracting the underlying continuum emission, which is dominated by stochastically heated dust grains and starlight from old stellar populations. 

A significant portion of our discussion focused on the challenges of developing PAH continuum subtraction prescriptions that can be robustly applied to a wide range of galaxies. There are a handful of approaches in the literature that describe continuum subtraction for the F335M, F770W, and/or F1130W filters \citep{Lai2020, Sandstrom2023, Bolatto2024, Chown2025, Whitcomb2025, Tarantino2025}, with more efforts in preparation. Our discussion concluded that it is vital to systematically compare these methods, coordinate across groups, and determine which approach to use in a given environment. Spectroscopic observations with JWST in a broader range of environments will be key to calibrate and test these methods. We also discussed that it is important to extend the nearby universe calibrations to higher redshifts, where the PAH bands will shift into different filters \citep[e.g.,][]{Lai2020}. 

A related topic of discussion was whether commonly used band ratios (e.g., 3.3/11.3 as a tracer of PAH grain size distribution) accurately reflect intrinsic PAH properties. Overall, we concluded that relative variations in these ratios are robust, but that using them to infer absolute quantities, such as the PAH ionization fraction, is unlikely to be reliable. We also noted that the choice of model, for example the PAH spectral database \citep{Ricca2026} vs. the \citet{Draine2021} models, can influence the inferred results.

\section{Need for more laboratory, models, and simulation work}
A clear theme of our discussion was the importance of laboratory data and simulations for interpreting PAH emission. In particular, we discussed that shorter wavelength features, such as the 3.4 $\mu$m aliphatic feature, are not well incorporated into current models, in part due to the lack of constraining observations during the \textit{Spitzer} era, which did not provide spectral coverage below 5 $\mu$m. With the launch of JWST and SPHEREx, these shorter wavelength PAH features are now receiving renewed attention. For JWST, the 3.3 $\mu$m PAH feature provides the highest spatial resolution tracer of PAH emission locally and is the last PAH feature to redshift out of JWST’s wavelength range. Further, SPHEREx will enable a local, all sky map of the 3.3 $\mu$m feature. Together, these capabilities present a growing opportunity to develop improved models of PAH emission at wavelengths below 5 $\mu$m.

Another avenue that simulations, models, and laboratory work can improve upon is our understanding of the PAH lifecycle. We discussed the direct detection of the rotational lines of two PAHs from TMC-1 \citep{McGuire2021} and how these results imply some amount of ``bottom-up'' formation, where PAHs are built in-situ in molecular clouds. There was some debate as to whether the physical conditions of TMC-1 can be broadly applied to all areas of astrophysics, as it is a particularly dense and cold cloud. Future simulations of the ``bottom-up'' vs ``top-down'' formation mechanisms for PAHs will be useful to compare with observations and determine the origin of PAHs.

\section{Problematic Acronym for Hydrocarbons}
In our discussion, we acknowledged that the carriers of the mid-infrared features may be more complex than simple PAH molecules (see Session 3, “What is in a name?” for more information). However, there was little enthusiasm for changing the terminology, as introducing yet another standard could lead to confusion, especially without widespread adoption\footnote{\url{https://xkcd.com/927/}}. Instead, we encouraged members of the PAH community to explicitly recognize that the carriers may be more complex than often assumed, recalling that PAHs have been described as a “Problematic Acronym for Hydrocarbons” (Greg Sloan).

\printbibliography[heading=subbibliography]
\end{refsection}

%% file: Structure/Session3.tex
\begin{refsection}
\chapter{What's in a name? Nomenclature of dust species / molecules}

\textbf{Moderators: Meriem Elyajouri, Karl Gordon}

\section{Context and motivation}
The title of this discussion session, ``What's in a name?'', was borrowed from Shakespeare's Romeo and Juliet, where Juliet argues that names are arbitrary labels: ``A rose by any other name would smell as sweet.'' In interstellar dust research, however, names carry significant weight. The terminology we use to describe carbonaceous species and their spectral signatures shapes how we conceptualize physical processes, which theoretical frameworks we apply, how effectively we communicate across sub-disciplines, and even which research directions attract funding.

This discussion was motivated by the recognition that the dust community uses overlapping and sometimes contradictory terms. This was evident throughout the conference itself. Depending on the speaker's background, the same emission features in the 3–20 $\mu$m range were referred to as \emph{PAH features}, \emph{Aromatic Infrared Bands} (AIBs), \emph{Unidentified Infrared Emission} (UIE), or \emph{carbonaceous features}. Similarly, a variety of carriers have been proposed: Polycyclic Aromatic Hydrocarbons (PAHs), mixed aromatic/aliphatic organic nanoparticles (MAONs), hydrogenated amorphous carbon (HAC or a-C:H), fullerenes, and more. 
Each term carries implicit assumptions about the physical and chemical nature of the emitting material. The distinction between ``large molecules'' and ``nano-grains'' also remains ambiguous, with different communities drawing the boundary at different sizes or not at all.

The timing is particularly relevant: JWST now provides unprecedented spectral quality and coverage of these features \citep[e.g.,][]{Peeters2024, Chown2024}, revealing substructure that challenges simplistic interpretations. Furthermore, different dust models, such as the Astrodust+PAH model \citep{Hensley2023} and the THEMIS model \citep{Jones2017}, attribute the observed features to fundamentally different carriers (isolated PAH molecules versus amorphous hydrogenated carbon nano-grains). The terminology we adopt can inadvertently bias interpretation.

This session addressed four themes: (1) the molecule--grain boundary, (2) whether PAHs should be classified as molecules, (3) terminology for mid-infrared features, and (4) paths forward. The goal was not to achieve consensus, but to articulate challenges and identify open questions for future consideration.

\section{Theme I: The molecule--grain boundary}
The first topic asked: is there a meaningful physical distinction between large molecules and small dust grains? At first glance, the distinction seems intuitive. Small molecules like H$_2$O are governed by quantum chemistry, while micron-sized particles exhibit solid-state behavior. However, the intermediate regime, tens to hundreds of atoms, spans three orders of magnitude in size. The group reached consensus that \textbf{no sharp dividing line exists}; the transition is continuous. Nevertheless, useful physical criteria were identified: \textbf{molecules} have discrete quantized energy levels and are treated with molecular physics, while \textbf{grains} are held together by van der Waals forces and exhibit collective solid-state behavior.

A crucial point emerged from dust models: the smallest nano-grains have radii of $\sim$3.5~\AA\ ($\sim$30 carbon atoms). Are such particles genuinely ``grains'', or better described as large molecules? At 30 atoms, discrete orbitals have not merged into bands, and behavior may be dominated by surface effects rather than bulk properties. The discussion also noted that the intermediate regime includes amorphous structures that may not fit neatly into either category.

\section{Theme II: Are PAHs really PAHs?}
From a chemical standpoint, \textbf{PAHs are unambiguously molecules}: fused aromatic rings with covalent bonds and discrete energy states. However, complications arise beyond idealized structures:
\begin{itemize}
    \item \textbf{Size limits}: Very large PAHs resemble graphene fragments. The molecule-to-grain transition remains unclear.
    \item \textbf{Structural modifications}: PAHs with sp$^3$ carbons, aliphatic side chains, or heteroatoms are no longer strictly PAHs chemically, though they \textit{remain molecules}.
    \item \textbf{Terminological conflation}: 
    Astrophysical literature often uses ``PAH'' to describe any small carbonaceous species emitting in the mid-IR, even when the carriers may include non-aromatic or amorphous material that would not qualify as PAHs in a strict chemical sense.
\end{itemize}



This matters because different dust models often use the same label, PAH, for physically different entities. The NASA Ames PAH IR Spectroscopic Database \citep{Bauschlicher2010, Boersma2014, Bauschlicher2018} treats PAHs as what they chemically are: individual gas-phase molecules with discrete electronic states, whose band strengths are computed from first-principles quantum chemistry (density functional theory) for hundreds of specific species \citep[NC =22--216;][]{Maragkoudakis2020}. The Draine models \citep[e.g.;][]{DraineLi2007, Hensley2023} also call their smallest carbonaceous component PAHs, but in practice treat them as nanoparticles characterized by a bulk carbon mass density, with band positions empirically tuned to match astronomical observations and band shapes described by Lorentzian profiles (not derived from the properties of any specific molecule). THEMIS \citep{Jones2017} goes a step further and avoids the term ``PAH'' altogether, yet the mid-IR signatures of its a-C(:H) nano-grains arise from the same kinds of aromatic and olefinic units that other models would label as PAHs, now treated as substructures embedded within a larger amorphous carbon network \citep{Jones2012c,JonesYsard2025}. In this picture, the infrared bands are only observable because these moieties are stabilized inside a more robust solid-state host nano-particle, rather than existing as isolated species in the gas phase. The result is that ``PAHs'' can refer to gas-phase molecules, to solid-like aromatic nanoparticles, or disappear entirely in favor of a substructure within an a‑C(:H) nano-grain, even though all of these formalisms are used to fit the same observational bands. This diversity is not just semantic, it highlights how our terminology can mask genuine physical differences between models while giving the misleading impression that we are all talking about the same carriers.  At the same time, this situation is perhaps not surprising. Modeling interstellar dust necessarily relies on a wide variety of carbonaceous materials (graphite, amorphous carbon, hydrogenated amorphous carbon, PAHs, fullerenes), and many different models reproduce the same observational constraints using different mixtures and parameterizations of these components \citep[][]{Siebenmorgen1992,Dwek1997,DraineLi2001,Zubko2004,Compiegne2011,Siebenmorgen2014,Jones2017,Draine2021}. The fact that multiple dust mixtures can fit the same data equally well \citep[15 satisfactory solutions in][alone]{Zubko2004} reveals a fundamental degeneracy and underscores how much remains unknown about the true nature of interstellar carbonaceous matter. In the face of such degeneracy, it is only natural that different communities have developed different conventions. But recognizing this is precisely why the discussion matters.

\section{Theme III: Terminology for mid-infrared features}
With JWST revealing the 3.3--20~$\mu$m features in unprecedented detail, nomenclature has become pressing. Participants reported using diverse terms: ``PAH features'' (common in North America), ``AIBs'' (emphasizing aromatic character), ``UIE'' (carrier-agnostic but now less common), ``carbonaceous features'' (broad umbrella), and specific identifications like ``fullerene bands''.

The discussion identified concerns with carrier-specific terminology. ``PAH features'' is too restrictive: contributors likely include aromatic and aliphatic species, olefinic structures, fullerenes, deuterated variants, hetero-PAHs, and amorphous nanoparticles. ``AIBs'' excludes aliphatic contributions (e.g., the 3.4~$\mu$m feature). Crucially, \textbf{terminology shapes interpretation}: calling something a ``PAH band'' biases modeling toward PAH properties.

The discussion clarified often-confused chemical terms: \textbf{aromatic} (cyclic with resonance stabilization, producing the 3.3~$\mu$m feature), \textbf{aliphatic} (sp$^3$ chains/rings, producing the 3.4~$\mu$m feature), and \textbf{olefinic} (C=C double bonds outside aromatic rings). There was general agreement that \textbf{carrier-agnostic terminology is preferable} when carrier identity is uncertain.

\section{Theme IV: Paths forward}
Participants identified criteria for good nomenclature: accuracy (not implying unwarranted specificity), flexibility (accommodating discoveries), clarity (understandable across sub-disciplines), and consistency. The discussion favored \textbf{``carbonaceous features''} as an umbrella term: it identifies elemental composition without assuming molecular structure, allows subcategories (emission/absorption, aromatic/aliphatic/olefinic, specific identifications when secure), and avoids privileging one model. Limitations were noted: ``hydrocarbonaceous'' might be more precise but is cumbersome. Related terminology exists in materials science (graphene quantum dots, carbon nanodots, carbon nanoparticles), but direct adoption could cause confusion across communities. Practical challenges include \textbf{inertia} (``PAH features'' appears in thousands of papers), \textbf{lack of enforcement} (adoption depends on community consensus), and \textbf{genuine uncertainty} about carriers.

\section{Conclusions and open questions}
\begin{enumerate}
    \item \textbf{No sharp molecule--grain boundary exists.} The transition is continuous, though useful distinctions can be made based on discrete vs.\ continuous energy levels and chemical vs.\ van der Waals bonding. The smallest ``grains'' in models ($\sim$30 atoms) may be better described as large molecules.
    \item \textbf{PAHs are molecules, but ``PAH'' is often misused.} Chemically defined PAHs are molecules. However, modified structures (with sp$^3$ carbons, side chains, heteroatoms) are no longer PAHs even if they remain molecules, and astrophysical usage is broader than chemical definitions warrant.
    \item \textbf{Carrier-specific terminology is problematic.} Terms like ``PAH features'' or ``AIBs'' embed assumptions that may bias interpretation toward particular models. The near- and mid-IR features likely arise from diverse carbonaceous species.
   \item \textbf{Our discussion favored ``carbonaceous features'' as a general term}, as it acknowledges what we know (carbon-bearing carriers) without assuming what we do not (specific structure), while allowing subcategories when carrier properties are confirmed. However, in the subsequent summary session vote with all conference participants, the term ``PAH features'' prevailed.
    \item \textbf{The interstellar carbonaceous material is more complex than our terminology suggests.} Aromatics, aliphatics, olefinics, fullerenes, amorphous particles... this diversity defies simple categorization. Nomenclature should reflect complexity rather than obscure it.
\end{enumerate}
\subsection{Open questions}
\begin{enumerate}
    \item Can the community converge on standardized terminology (e.g., through IAU recommendations), or will multiple conventions persist?
    \item How should we handle the intermediate regime between molecular and solid-state physics in models?
    \item As JWST resolves new spectral substructures, how do we name features without prematurely assigning carriers?
    \item Can nomenclature be harmonized with materials science (carbon nanodots, graphene quantum dots)?
\end{enumerate}
The questions raised here will not be resolved quickly, but critically examining the terms we use to describe the interstellar material is valuable as observational and theoretical capabilities advance. We encourage the community to continue this conversation. The next dust meeting will find some questions answered and new ones will have emerged; recording our current thinking provides a reference point for future discussion.

\printbibliography[heading=subbibliography]
\end{refsection}

%% file: Structure/Session4.tex
\begin{refsection}
\chapter{The UV bump}

\textbf{Moderator: Irene Shivaei}\\
Discussion participants (alphabetical order): 
Monique Aller, B-G Andersson, Leonor Arriscado, Bruce Draine, Andrew Mizener, Kaylee Perez, Noel Richardson, Samir Salim, Irene Shivaei

\section{Motivation}
One of the most distinct features of the extinction/attenuation curves of galaxies is the broad absorption feature centered at 2175$\AA$, so called ``UV bump''. In the local Universe, the MW and LMC extinction curves both have strong UV bumps, but the bump is very weak or absent in the SMC curve \citep{gordon24}. The attenuation curves of galaxies (derived from integrated light of the galaxy, as opposed to single sight-lines) show similarly a large variation in the presence and strength of the UV bump: the bump is absent in the average starburst attenuation curve at $z\sim 0$ \citep{calzetti00}, and at higher redshifts there is a range of bumps with more massive and metal-rich ones showing, on average, stronger bumps \citep[e.g.,][]{noll09,kriek13,zafar18,shivaei20,battisti20,kashino21,shivaei22}. With JWST, now we are able to spectroscopically probe rest-frame UV of galaxies during the reionization epoch, and while rare, some very young galaxies show signs of a strong UV bump in their spectra \citep[e.g.,][]{witstok23, markov25, ormerod25}.

This session's discussion was focused on understanding the origin, variability, and observability of the UV bump in galaxies, with particular emphasis on its connection to dust properties and polycyclic aromatic hydrocarbons (PAHs). Key guiding questions included:

\begin{itemize}
    \item Why is the UV bump absent or weak in some galaxies?
    \item How is the UV bump related to PAH emission and carbonaceous dust?
    \item What is the role of radiative transfer, scattering, and geometry in shaping observed extinction and attenuation curves?
    \item Why is the central wavelength of the bump remarkably stable in the Milky Way, while its width varies?
    \item How do metallicity, environment, and cosmic time affect the presence and properties of the bump?
    \item Can related absorption features, such as diffuse interstellar bands (DIBs), help identify the carriers of the bump?
\end{itemize}

\section{Key observational and physical issues}

\subsection{Relation between PAHs and the UV bump}

A central theme was the expected correlation between PAH emission and the UV bump. In principle, regions rich in PAHs should also display a prominent bump if PAHs are its main carriers. More observations are needed to establish this correlation. However, current observations over smaller samples show cases where strong PAH emission is present without a clear UV bump. This challenges simple interpretations and suggests that either PAHs are not the sole carriers (small carbonaceous grains as alternative carriers), or that observational effects strongly modulate the bump appearance. The spatial distribution of PAHs and dust was therefore discussed as a path forward, with the recognition that PAHs can exist in a wide range of environments and are not confined to classical HII regions. 

\subsection{Extinction versus attenuation}

Participants emphasized that many extragalactic studies rely on attenuation curves, which are shaped not only by intrinsic dust properties but also by complex geometries and radiative transfer effects. Scattering and absorption can combine in ways that significantly weaken or even erase intrinsic spectral features. If the bump carriers are small particles with weak scattering (as expected for small PAHs), stronger correlations between PAH and bump strength might be expected. Conversely, if graphite-like grains dominate, scattering may play a major role. This uncertainty makes it difficult to interpret weak bumps as evidence for the absence of specific dust components. 
Therefore, radiative transfer and geometry were identified as sources of uncertainty. In unresolved galaxies, mixed star–dust configurations introduce significant noise and degeneracies. Even in nearby systems, separating direct, scattered, and absorbed light remains challenging.
As a potential way forward spatially resolved spectroscopy over nebulae, to study reflected and transmitted light, combined with modeling the 3D distribution, was suggested to help disentangle absorption and scattering contributions (do we see the bump in scattered light or only in the absorption in the center?). Additionally, studies of face-on or simple geometries (or extinction curves) can be promising.

\subsection{Technical fitting challenges}

A major technical difficulty that affects the measurement of the UV bump is the degeneracy between the far-UV rise and the bump strength. Sensitivity to the continuum placement and the assumed underlying extinction/attenuation curve make the bump measurements uncertain. Additionally, limited wavelength coverage and signal-to-noise, particularly at high redshifts, make the measurements more uncertain. These issues complicate comparisons across different samples and environments.

\subsection{Central wavelength and shape of the bump}

A major unresolved issue is the remarkable stability of the bump’s central wavelength in the Milky Way, despite significant variations in PAH band ratios. Despite significant variations in PAH band ratios, the peak position remains nearly constant, while the bump width shows measurable variations. This stability places strong constraints on theoretical models of the carrier. Participants questioned whether similar behavior is observed in nearby galaxies such as M31 and M33 and noted tentative evidence for shifts in the central wavelength at high redshift (cosmic noon and above). Determining whether these shifts are real and understanding their physical origin, and whether there are links between bump shape and PAH molecular structure, remain key priorities.

\subsection{Metallicity and environmental effects}

The presence and strength of the UV bump appear to correlate with metallicity, both locally and at high redshift. However, the nature and robustness of this correlation remain uncertain. There are other factors that can affect the bump strength and introduce scatter in the bump-metallicity relation, among these are: star formation intensity, radiation field hardness, dust processing and destruction, and evolution of carbonaceous dust over cosmic time. Studies with larger samples (both locally and at high redshifts) are needed to pave the way.

\subsection{High-redshift studies}

High-redshift observations show that UV bumps do exist at early cosmic times, indicating early formation of carbon-rich dust. However, the average attenuation curve at high redshift lacks a strong bump. This raises questions about selection effects and limited spectral resolution, mixing of diverse environments in integrated spectra, and the need for controlled, well-characterized samples. High-redshift studies were viewed as valuable probes of dust evolution and identification of carbon dust in young systems, but interpretations depend on robust local benchmarks.

\subsection{DIBs and laboratory studies}

The potential connection between the UV bump and diffuse interstellar bands (DIBs) was discussed. In an ideal scenario, identifying DIB carriers and correlating their strengths with the bump could reveal the nature of the responsible molecules. However, this goal remains challenging. Hundreds of DIBs are still unidentified, and laboratory experiments are extremely challenging, as large molecules must be studied in the gas phase and tend to adhere to experimental surfaces (and easily freeze out on the walls of chambers). The identification of $C^+_{60}$ as a carrier of specific DIBs was cited as a notable success, but it does not explain the full range of PAH-related features. Variations of DIBs and the UV bump in specific environments, such as around Wolf–Rayet stars and face-on systems, were highlighted as promising test cases.

\section{Key conclusions, challenges, and future directions}
From the discussion, several broad conclusions emerged. 

\begin{itemize}
\item The UV bump is almost certainly linked to carbonaceous material, but the dominant carrier of the UV bump (whether PAHs, small carbon grains, graphite-like particles, or a mixture) has yet to be identified. More observations with PAH and bump (in extinction or attenuation), particularly spatially resolved, can help. Expanding laboratory efforts on large carbonaceous molecules and building comparative samples in nearby galaxies were also seen as essential.

\item Radiative transfer and geometry influence observed attenuation curves and can mask intrinsic dust properties, therefore extinction observations are invaluable. Particularly, conducting spatially resolved spectroscopic surveys of nearby systems, and also combining extinction measurements from high-redshift background sources such as lensed GRBs and quasars with attenuation studies, can be very informative. Additionally, developing more realistic radiative transfer models, and systematically comparing bump properties and polarization measurements are paths forward.

\item Current observational limitations and fitting uncertainties remain a major obstacle, particularly when wavelength coverage and signal-to-noise ratio are limited. 

\item The physical mechanism behind the stability of the bump's central wavelength within MW sightlines is unknown, as is the universality of this behavior. Comparing PAH band ratios and bump properties may provide clues on the physical origin of the shape and central wavelength of the bump.

\item Finally, high-redshift detections indicate early carbon dust production, but assembling homogeneous samples across cosmic time is needed to bring forward a coherent picture on the significance and frequency of the bump across galaxy populations and redshifts. Local galaxies studies are essential to interpret the high-redshift findings.
\end{itemize}

Overall, the discussion highlighted the need for coordinated observational, theoretical, and laboratory efforts. Only by combining detailed local studies with high-redshift surveys and improved modeling will it be possible to fully understand the origin and evolution of the UV bump and its connection to interstellar properties.

\printbibliography[heading=subbibliography]
\end{refsection}

%% file: Structure/Session5.tex
\begin{refsection}
\chapter{Combining results from laboratory, theory and observations}

\textbf{Moderator: Thomas Henning}\\

In astronomy, one often finds relatively inaccurate descriptions of the material properties of cosmic dust, present in the diverse environments of interstellar and circumstellar space. However, optical properties of dust particles depend on their exact composition, particle size and shape, and even intrinsic dust temperature. Therefore, the use of accurate terminology is strongly recommended.

In the interstellar medium, silicates occur exclusively in amorphous form, with their optical properties depending on the composition (e.g., elemental Mg/Fe ratio) and structure of the material – the polymerization state. The polymerization state of amorphous silicates ranges from less cross-linked (more fluid/soluble materials) to highly polymerized amorphous silicates. Crystalline silicates include enstatite - the magnesium-rich end-member of pyroxenes - and forsterite – the magnesium-rich end- member of the olivine solid solution series. Such crystalline silicates have been discovered in planet-forming disks and comets, indicating high-temperature annealing and radial mixing.

Atoms in carbon-based solids can bond together in a variety of ways, leading to different allotropes of carbon. These carbon-based solids range from conducting graphitic material with sp2 hybridization with non-localized $\pi$ electrons to non-conducting diamond-like structures with bonding through sp3 hybridized orbitals. Other allotropes include the material family of fullerenes and graphene. Carbonaceous solids can occur both in the laboratory and in space as non-crystalline solids. The amorphous material can be stabilized by dangling $\pi$ bonds with hydrogen atoms. The material properties depend on the ratio of sp2 to sp3 hybridized bonds, the arrangement of basic structural units, and additional structural parameters. On a more macroscopic level, the optical properties will depend on particle size and shape, with the properties of nanoparticles deviating from the properties of bulk materials.

Both in laboratory studies of the optical properties of dust and in modeling efforts the material properties should be characterized as comprehensively as possible so that the results of different studies can be compared. Modeling results may strongly depend on the assumed optical dust properties. These properties should be treated as additional ``parameters'' in the models, especially if a strong spatial or temporal variation of the dust properties can be expected. An example is the derivation of dust masses from submillimetre/millimetre observations where a specific dust opacity has to be assumed. In this context, we should note that there is no single material which has a constant $\beta$ value for the wavelength dependence of the absorption efficiency from mid-infrared to millimetre wavelengths and that the wavelength dependence for specific materials can depend on the dust temperature.

For many materials optical constants have been measured in the laboratory, but often only for a limited wavelength range and data need to be extrapolated for the application in radiative transfer simulations. For certain classes of materials new data are needed (e.g. carbides), and the transition from nano-sized materials to bulk materials has to be better described and understood. There is a plea to reference the applied experimental data in modeling papers and not just refer to another modeling paper where the same data have been used.

The solid reservoirs of dust components such as iron and sulfur remain to be better understood. Iron is heavily depleted in the diffuse interstellar medium and probably accreted by other dust components. Iron is certainly providing an important opacity source for silicates in planet-forming disks, leading to the dust temperature derived from infrared silicate emission profiles. Sulfur becomes depleted in molecular cloud environments and must be an important ice component, possibly as complex organo-sulfur molecules, but potentially also in iron sulfides.  For the molecular ices, the interaction between the solid surfaces and the molecular ice components must be better studied, especially the potentially catalytic processes triggered at this interface. The optical properties of mixed ices is an important research topic for the future.

So far, most of the dust formation and destruction models for galaxies indicate that dust production through re-formation by accretion of atoms in the interstellar medium is an important grain formation process in the local universe. Indeed, laboratory experiments have demonstrated that separate components of amorphous silicates and carbonaceous dust can be produced in low-temperature reactions. Dust formation in the interstellar medium does not necessary lead to inhomogeneous dust (mixture of silicates and carbonaceous material). The net dust formation efficiency in supernovae remains an open question as well as the contribution of AGB stars at higher redshift to the dust production. In this context it should be noted that PAHs can be formed by a large variety of processes from high-temperature combustion-type reactions to low-temperature formation pathways and finally through shocks where PAHs sitting on dust surfaces can be ablated. Galaxy-wide studies of PAH abundances will help to unravel the most important production and destruction processes of PAHs.

The information content of infrared and (sub)millimetre dust observations both from the ground and space can only be unlocked if combined efforts of laboratory astrophysics groups and modeling of dust formation, evolution, and destruction are supported together with the observational facilities.

\printbibliography[heading=subbibliography]
\end{refsection}

%% file: Structure/Session6.tex
\begin{refsection}
\chapter{Priorities for incorporating dust in simulations}

\textbf{Moderator: Maarten Baes}\\
\textbf{Additional authors: John-David Smith, Desika Narayanan, Caleb Choban}

\section{How important is dust in cosmological simulations?}

Dust plays a central role in galaxy formation and evolution, influencing star formation, the thermodynamic state of the interstellar medium (ISM), and the interpretation of galaxy observations across the electromagnetic spectrum. Despite representing only a small fraction of the baryonic mass budget, dust regulates radiative transfer, catalyses molecule formation, and shapes the relationship between intrinsic and observed galaxy properties. The discussion converged on a shared view: dust is no longer a secondary or optional post-processing detail, but should be a core component of the next generation of galaxy formation models. And indeed, modern cosmological large-volume and zoom simulations increasingly include dust as an explicitly evolved component, tracking its production, growth, destruction, and transport alongside gas, stars, and metals \citep[e.g.][]{McKinnon2017, Aoyama2018, Aoyama2020, Li2019, Graziani2020, Granato2021, RagoneFigueroa2024, Han2025, Schaye2025}. 

However, all such treatments rely on subgrid prescriptions. Dust formation in AGB winds and supernova ejecta, accretion-driven grain growth in dense gas, destruction in supernova shocks, thermal sputtering in hot gas, and grain--grain collisions all occur on scales far below those resolved in cosmological simulations. As a result, dust evolution models are necessarily parameterised \citep[e.g.][]{Dwek1998, Asano2013, Hirashita2015, Choban2022, Narayanan2023, Matsumoto2024, Dubois2024, Trayford2025}. Many of the detailed parameters of these models are still very uncertain. Galaxy simulations with dust should therefore be viewed as numerical experiments: their strength lies not in providing a fundamental theory of dust, but in testing how different physical assumptions manifest in galaxy-scale observables. This framing is essential for interpreting both successes and tensions between simulations and data.

\section{How can simulations help us understand the physics of dust?}

Simulations are not only tools for accommodating dust, but also for learning which dust processes dominate under different conditions. Achieving this requires a combination of bottom-up and top-down approaches.

On small scales, detailed models of dust formation and processing provide critical input. These include dust condensation in AGB winds \citep{FerrarottiGail2006, Nanni2013, Ventura2018, Ventura2020}, dust formation and survival in supernova ejecta and remnants \citep{Nozawa2007, Bocchio2016, Scheffler2025}, and grain destruction in shocks and hot gas \citep{Tsai1995, Farber2022}. Such studies inform yields, condensation efficiencies, size distributions, and destruction efficiencies, many of which remain uncertain at the order-of-magnitude level.

On larger scales, cosmological and zoom simulations embed these processes in evolving galactic environments. A consistent result emerging from both analytic arguments and numerical models is that grain growth in the ISM becomes the dominant contributor to the dust mass once galaxies reach moderate metallicities; at earlier times, however, dust evolution is highly sensitive to assumptions about initial grain sizes, shattering efficiencies, and destruction rates \citep[e.g.][]{Asano2013, Zhukovska2014, McKinnon2017, Graziani2020, Trayford2025}.

\section{Which observations best constrain dust in simulations?}

Polycyclic aromatic hydrocarbons (PAHs) were highlighted as a ``gift of nature.'' Their prominent mid-infrared emission features provide strong constraints on the abundance of small grains, local radiation fields, and environmental processing \citep{Draine2007, Galliano2018}. As emphasised by \citet{Narayanan2023} and \citet{Matsumoto2024}, PAHs are especially powerful because they respond rapidly to changes in ISM conditions, making them sensitive probes of dust evolution models.

Far-infrared and submillimeter observations are key to mapping the bulk of the dust content in galaxies. Spatially resolved imaging at multiple far-infrared bands is essential to disentangle the spatial and temperature distribution of the dust. In light of the forthcoming PRIMA mission \citep{Glenn2025}, polarimetric far-infrared observations were identified as a particularly interesting observational constraint. These observations probe grain shapes, alignment, and magnetic fields, offering information inaccessible to total-intensity measurements alone. Incorporating such constraints into simulation-based forward models remains challenging, but the potential diagnostic power is substantial \citep[e.g.][]{Vandenbroucke2021}.

Finally, UV extinction curves were emphasised as a key observable for constraining intrinsic dust properties. Unlike attenuation curves, which depend strongly on geometry and radiative transfer effects \citep[e.g.][]{WittGordon2000, SeonDraine2016, SalimNarayanan2020}, extinction curves are more directly tied to grain size distributions and composition. Features such as the far-UV slope and the 2175~\AA\ bump provide particularly strong constraints, and recent simulations demonstrate that processes like shattering and grain growth leave clear imprints on their evolution \citep[e.g.][]{Li2020, Matsumoto2024}.

\section{Looking forward}

In summary, the discussion underscored that dust is both indispensable and challenging to model in galaxy simulations. Progress will require tighter integration between small-scale studies of dust microphysics, large-scale galaxy simulations, and targeted observations. Priorities include improved subgrid prescriptions informed by detailed modelling, broader adoption of grain-size-dependent dust models, and forward modelling of high-information observables such as PAH emission, polarisation, and extinction curves. With these advances, simulations will increasingly be able not only to include dust realistically, but also to use dust as a diagnostic tool for understanding galaxy evolution across cosmic time.

\printbibliography[heading=subbibliography]
\end{refsection}

%% file: Structure/Session7.tex
\begin{refsection}
\chapter{Evolution of dust properties in different environments and physical conditions}

\textbf{Moderator: Andrew Saydjari}

\section{Motivation}
There are two key motivations for understanding the evolution of dust properties: (1) to leverage the variation of dust properties as a tracer for variations in the physical conditions of the ISM and (2) so we can better correct for the impact of dust on observations. The first is more difficult, and requires connecting physical processes in the ISM with their impact on the observational properties of dust grains. This spurred lots of exciting discussion on connections between theory, laboratory experiment, and observations. The second is a predominantly observational question, which spurred a discussion of important measurements that are upcoming or that we should be designing to probe dust property variations.

\section{Do we need a biography for each dust grain?}

One thematic question for the session was trying to understand the difficulties involved in connecting current and past physical conditions for dust grains with their current properties... and the regimes where this difficult task might be possible. The challenge was nicely summarized by the question ``Do we need a biography for each grain?'' Dust grains experiencing the same current physical conditions may still have different properties if they have different histories. This requires at least some handle on the time-dependent conditions during the lifetime of the grain, which might be possible on average for ensembles of grains by combining upcoming kinematic dust maps with simulations. Presolar grains studied in the lab provide a special case where writing a dust grain biography may be possible. However, many were concerned about the biased sample these present and how the properties and histories of grains that were destroyed differ. Going beyond questions of dust grain lifetimes to answer the question ``How long do dust grains remember where they came from?'' will be necessary to understand if efforts to study dust properties as a function of distance from plausible production centers is feasible. To that end, laboratory experiments on carbon-silicon atom sticking and the question of dust grain compositional homogeneity were discussed.

\section{What tracers do we want (more of)?}

While we \emph{always} want more observations, discussion focused on (1) which dust grain and environmental properties were least constrained and (2) important combinations of different observations. Of particular interest were large-area measurements of dust grain porosity, which would require the development of polarization survey(s). Ways to overcome difficulties with degeneracies between porosity and grain aspect ratio as well as the benefits of comparing both optical and FIR polarization were discussed, especially in the context of PRIMA. Spectropolarimetry of specific dust extinction curve or emission features (such as the 3.4 $\mu$m PAH band) would also be extremely elucidating.

A broad spectrum of correlative studies were called for between UV radiation field strength, metallicity, dust-to-gas ratio, PAH feature strength, and gas-phase depletion. One concrete example of an understudied bridge is studying correlations between dust extinction curve variation and gas-phase depletion measurements. While large catalogs of dust extinction curve variation measurements are available, doing so would require significantly expanding the number of gas-phase depletion measurements. We also discussed recent exciting progress on combining 3D dust maps with stellar catalogs to produce maps of the UV radiation field and other tracers like H$\alpha$. Concerns about the treatment of dense cloud cores and more general concerns about degeneracies between changes in dust properties and radiation field variations were raised. Careful theoretical work and temperature measurements may help in disentangling some of these effects, such as shielding and the deposition of molecules onto grains.

\printbibliography[heading=subbibliography]
\end{refsection}

%% file: Structure/Session8.tex
\begin{refsection}
\chapter{Dust and galaxy evolution}

\textbf{Moderator: Chian-Chou Chen}\\

This discussion session focused on two broad topics related to dust in galaxies across cosmic time. The first addressed the connection between dust and star formation history, followed by a discussion on the determination of dust masses. The details of each topic are summarized below.

\section{Dust and star formation history}
Motivated by recent studies of the early Universe, the discussion examined how a galaxy’s star formation history may influence the properties of its dust. The specific dust properties under consideration — such as total mass, composition, and grain size — did not come across as clearly defined during the discussion. However, any potential connection between dust and star formation history was generally thought to be mediated by the interaction between supernovae and dust, through both dust production and destruction.

However, based on the current state of knowledge and as emphasized in the review talks at the conference, dust destruction by supernovae remains a major and largely unconstrained uncertainty. Although dust production mechanisms are relatively better understood, the dominant sites of dust growth and the detailed physical processes involved remain poorly constrained. As a result of these uncertainties, the discussion concluded that it is currently unclear whether a robust and direct relation between star formation history and dust properties can be established.

\section{Determination of dust mass}
Dust mass in the interstellar medium is a fundamental quantity in studies of galaxy formation and evolution. Measurements of dust masses in uniformly selected galaxy samples enable the construction of dust mass functions, which provide key constraints for theoretical models. Moreover, mass ratios among dust, gas, stars, and metals offer critical insights into how galaxies assemble and evolve across cosmic time. Recognizing the importance of these measurements, the determination of dust mass, and an assessment of its associated systematics, was one central focus of this discussion session.

Regardless of the method used to estimate dust masses, whether through modified blackbody (mBB) fitting or energy-balance spectral energy distribution (SED) modeling, one major source of systematic uncertainty agreed upon by the participants is the dust mass absorption coefficient, $\kappa_\nu$. This uncertainty is commonly believed to be on the order of a factor of $\sim$2–4. The uncertainty associated with $\kappa_\nu$ ultimately reflects our limited understanding of the physical properties of interstellar dust. In particular, $\kappa_\nu$ depends sensitively on dust grain composition and size distribution, and other factors, all of which may vary significantly between different galactic environments. Processes such as grain growth through coagulation in dense regions, shattering in diffuse media, or rapid chemical enrichment at high redshift can alter the effective opacity per unit dust mass. 

In addition, the simplified treatment of dust geometry in mBB fitting, which normally assumes a foreground screen of dust, could introduce further uncertainty, potentially adding another factor of $\sim$2–4. Taken together, these effects can result in total systematic uncertainties of $\sim$5–10 in dust mass estimates from mBB fitting, with galaxies exhibiting extreme dust geometries, such as compact starbursts, being particularly affected.

By contrast, when sufficiently broad wavelength coverage is available (from ultraviolet to far-infrared), energy-balance SED modeling is generally recommended. This approach provides some leverage on the dust geometry, although uncertainties remain due in part to variations in attenuation curves. Under these conditions, the total systematic uncertainty in dust mass estimates can likely be reduced to a factor of $\sim$3–5.

These uncertainties have important implications for studies that compare dust masses across redshift or environment. Applying locally calibrated values of $\kappa_\nu$ to high-redshift galaxies implicitly assumes minimal evolution in dust grain properties, an assumption that remains largely untested. Consequently, apparent trends in dust mass functions or dust-to-stellar ratios with cosmic time may be influenced, at least in part, by systematic effects rather than genuine physical evolution. Care must therefore be taken when interpreting differences between samples analyzed with different methodologies or assumptions, particularly when such differences are at the level of a factor of a few.

Ultimately, the dominant and irreducible uncertainty in dust mass measurements remains that associated with $\kappa_\nu$. Looking ahead, we recommend that studies explicitly state their assumptions regarding this parameter and other key systematics, and that such uncertainties be carefully accounted for when comparing results across different analyses.

\printbibliography[heading=subbibliography]
\end{refsection}

%% file: Structure/Session9.tex
\begin{refsection}
\chapter{Connecting Solar System observations to ISM dust}

\textbf{Moderator: Aki Takigawa}

\section{Primitive information preserved in Solar System materials}
During the formation of the Solar System, the majority of dust in the protoplanetary disk accreted onto the Sun, while a portion was incorporated into planetesimals to become the building blocks of planets and small bodies. Among these, planetesimals that have retained the isotopic composition and mineralogical information of the protosolar disk dust until the present day, having undergone minimal thermal metamorphism or aqueous alteration, are defined as the ``most primitive bodies'' in the Solar System. Specifically, comets and the parent bodies of least altered carbonaceous chondrites fall into this category. Therefore, the detailed analysis of cometary dust and primitive chondrites is crucial for linking current Solar System materials with the molecular cloud material from which our Solar System originated.
\section{The discrepancy between presolar grains and circumstellar dust}
Presolar grains are identified as survivors of circumstellar dust that existed prior to Solar System formation, as their extreme isotopic anomalies match theoretical values of stellar nucleosynthesis. Generally, primitive bodies contain higher abundances of presolar grains; however, even in cometary dust, which is considered the most primitive, their abundance is estimated to be less than 1\% \citep{Floss2016}.
In other words, the vast majority of cometary dust does not exhibit large isotopic anomalies. Two possibilities can explain this:
\begin{itemize}
     \item Homogenization of Interstellar Dust: The interstellar dust flowing into the Solar System did not possess isotopic anomalies to begin with.
     \item  Loss in the Disk: Most dust lost its anomalies due to isotope exchange reactions during the Solar System formation process.
\end{itemize}	
The first possibility is consistent with the astronomical models that over 90\% of interstellar dust originates from dust re-formation (condensation and growth) in the interstellar medium \citep[e.g.,][]{Draine2009}. Dust re-formed in interstellar space is expected to be homogenized, averaging out the isotopic anomalies of individual stars.
Regarding the second possibility, isotope exchange experiments on amorphous silicates have shown that isotopic anomalies in the amorphous phase can be lost within the lifetime of the disk \citep{Yamamoto2018}. In contrast, crystalline silicates undergo isotope exchange much more slowly, resulting in a higher probability of survival as presolar grains. Indeed, the fact that a few tens \% of presolar silicates are  crystalline, whereas observed circumstellar dust is predominantly amorphous, is consistent with this ``selective loss'' scenario.
Furthermore, while metastable alumina is frequently observed in variable stars \citep{Sloan2003}, it is rare among presolar grains \citep{Stroud2004}, which also suggests thermal and chemical processing within the disk.\\
Care must be taken regarding the ``baseline'' used when comparing interstellar dust with Solar System materials. If interstellar dust is formed via re-formation in the interstellar medium, nucleosynthetic anomalies are likely homogenized. However, ``homogeneous'' in this context implies a composition close to the proto Sun, which differs from the terrestrial standard.
The GENESIS mission estimated the oxygen isotopic composition of the Sun to be approximately $\delta^{18}\mathrm{O} \approx -59$\textperthousand, which is isotopically lighter than the value for Earth and meteorites ($\approx 0$\textperthousand). This discrepancy can be explained by the mixing and evolution of solar-composition gas with heavy oxygen (fixed as ice following CO dissociation) resulting from self-shielding effects in the molecular cloud \citep{Yurimoto2004}.
Therefore, explaining the current oxygen isotopic composition of the Solar System requires a scenario where ``interstellar re-formed material with a solar-like (light) isotopic composition evolved into an Earth-like (heavy) composition within the solar disk.''
\section{GEMS: Survivors of re-formed interstellar dust?}
Cometary dust is rich in particles known as GEMS (Glass with Embedded Metal and Sulfides), which are amorphous silicates containing metal and sulfide nanoparticles. The origin of GEMS has been debated, with theories proposing they are either ``survivors of irradiated circumstellar dust'' or ``condensates formed within the Solar System'' \citep[e.g., ][]{Keller2011,Bradley2013}.
However, if we adopt the view that the majority of interstellar dust is re-formed in the interstellar medium, a third possibility emerges: Most GEMS are re-formed interstellar dust grains themselves.
Based on this hypothesis, GEMS originally possessed a solar-like light oxygen isotopic composition. Through low-temperature processes in the disk, they underwent oxygen isotope exchange without crystallization, shifting to the current Earth-like composition. In this case, cometary dust, which has the highest abundance of presolar grains yet shows no isotopic anomalies in its bulk majority, can be identified as the actual ``processed interstellar dust.''
However, it should be noted that the average chemical composition of GEMS is Si-rich relative to Solar System elemental abundances. Whether such a composition can be fully explained by dust formation models in the interstellar medium (ISM) remains a subject for future investigation. Furthermore, the microstructural features, specifically Fe nanoparticles embedded within the amorphous silicates and Fe sulfide grains on the grain surfaces, require distinct explanations. The Fe sulfide grains likely formed via sulfidation in the protosolar disk. In contrast, while the direct condensation of metallic Fe nanoparticles in the ISM has not been experimentally demonstrated, low-energy ion-dust interactions in the ISM have been proposed as a plausible mechanism for forming such structures \citep{RN648}.

\section{Evolution of organic matter and future prospects}

The organic matter co-accreted with dust reflects a complex evolution. While the diversity of soluble organic matter (SOM) in Ryugu and carbonaceous chondrites is often attributed to parent-body aqueous alteration, recent studies suggest that their precursors originated from reactions in the interstellar medium (ISM). Thus, SOM is a product of continuous evolution from the ISM to the parent body. Similarly, the ``primordiality'' of insoluble organic matter (IOM) must be carefully evaluated. Although it is often cited as a reservoir of molecular cloud signatures, IOM is susceptible to thermal and radiation processing in the disk, which can overprint its original characteristics. 
Furthermore, explaining the carbon depletion in the inner Solar System \citep{Binkert2023} requires a better understanding of how these organics were transformed during disk transport. Future research must bridge the gap between astronomical observations and sample analysis through integrated models.\\
Finally, it must be noted that the cometary samples currently available are limited to interplanetary dust particles (IDPs) and samples from the coma of Comet Wild 2 returned by the Stardust mission. To verify these arguments and unravel the full picture of the evolution from interstellar matter to planetary systems, future sample return missions from cometary nuclei and in-situ analysis missions like Comet Interceptor will play a decisive role.

\printbibliography[heading=subbibliography]
\end{refsection}

%% file: Structure/Session10.tex
\chapter{Dust destruction}

This session was canceled.

%% file: Structure/Session11.tex
\begin{refsection}
\chapter{Supernova dust properties}

\textbf{Moderator: Jacob Jencson}

\section{What can we learn about dust formation from JWST observations of extragalactic supernovae?}

With the launch of the James Webb Space Telescope (JWST), the sample of extragalactic supernovae (SNe) for which we can attempt to trace the formation of dust grains in the mid-infrared (IR) is growing rapidly \citep[e.g.,][]{shahbandeh23}. The discussion identified a number of challenges in interpreting these observations: (1) mid-IR spectral energy distributions are often sparsely sampled both in wavelength and time, with significant ambiguities in the origin and location of the emitting warm dust (i.e., pre-existing circumstellar dust vs.\ newly formed ejecta dust; see, e.g., \citealp{pearson25}), (2) the inclusion of cooling in the interpretation of the observed gradual ($\sim$20~year) dust mass trend, as only a small fraction of the total dust mass ($\sim$0.1\%) may be warm enough to contribute to the mid-IR flux, (3) while the mid-IR has features useful for constraining dust compositions and the presence of molecules or PAHs, the effects of optical depth complicate this picture, (4) moreover, far-IR observations could constrain the total cold dust masses but lack features sensitive to composition for the bulk of the dust mass. Despite these challenges, the discussion participants emphasized the importance of obtaining detailed, time-resolved spectral sequences for at least a small subset of SNe that track the evolution of dust features together with precursor molecules (e.g., CO) to help discriminate between slow grain growth and scenarios involving optically thick clumps that could suppress mid-infrared signatures.

The group also discussed the importance of considering other types of transients as contributors to the cosmic dust budget, including the so-called ``gap transients'' like stellar mergers and SN impostors, as well as classical novae. The question of the contributions of SNe~Ia, despite their higher ejecta velocities and lack of evidence for dust formation in known remnants, was also raised in relation to the question of Fe-depletion in the interstellar medium.  

\section{How can we utilize exquisite observations of dust in SN remnants?}
Galactic SN remnants provide a far more detailed view of supernova dust, now with exquisite data on individual dust clumps that have revealed strong spatial and compositional variations. However, interpreting these data remains challenging, as spectral models often rely on a handful of simplified dust components and uncertain optical constants. Persistent discrepancies include the prominence of unexpected species (e.g., SiO$_2$), and difficulties in securely identifying or excluding components like graphite due to a lack of clear features. Participants indicated areas where improved modeling could enable progress related to the structure and composition of progenitor stars, explosion models, grain nucleation (especially for small $~\sim$10-atom grains), and particularly the level of mixing in the ejecta. The group also discussed the need for improved, physically motivated sets of optical constants and specifically highlighted that laboratory measurements derived from actual pre-solar grains of supernova origin could provide more realistic constraints for dust models.

\section{How can we make progress understanding dust destruction by the reverse shock?}

Estimates of the fraction of newly formed supernova dust that will be destroyed by the passage of the reverse shock have spanned essentially the full range from 0--100\%. A major limitation of current modeling is often the simplicity of the circumstellar environment in which the shock propagates. In reality, SN progenitors likely have complicated circumstellar density profiles with significant asymmetries arising from complex mass-loss histories involving episodic mass ejections, binary interactions, and pre-SN outbursts. SN remnants, for example, demonstrate the presence of circumstellar rings, disk-like geometries, and large density variations. The ultimate level of heating and destruction of dust grains is heavily dependent on the interaction of the reverse shock with such complicated circumstellar environments. SN remnants like Cas A, where we have huge amounts of information on these interactions, will be especially useful in answering questions related to dust destruction in SNe. 

\printbibliography[heading=subbibliography]
\end{refsection}

%% file: Structure/Session12.tex
\begin{refsection}
\chapter{Dust and PAHs in AGN environments}

\textbf{Moderator: Thomas Lai}

\section{Motivation and community interests}
JWST has opened a unique opportunity to investigate dust and PAHs in AGN environments with unprecedented spectral and spatial resolution. PAHs have been considered potentially sensitive diagnostics of ISM conditions and are widely used as tracers of star formation rate (SFR). However, the relationship between PAH emission and SFR can break down in the presence of an AGN. These applications critically depend on our understanding of PAH properties and how they vary with metallicity and respond to ambient radiation fields. PAH emission can be weak or even absent in low-metallicity systems, as well as in galaxies hosting powerful AGN. Surprisingly, some studies have shown that AGN photons can also be the source of PAH molecule excitation and emission, further complicating the interpretation.

In this discussion session, we selected several predefined topics relevant to dust and PAHs in AGN and ranked them based on participant interest. The resulting ordering provides a snapshot of the community’s current priorities in understanding the interplay between AGN activity and dust properties. The topics discussed, in descending order of interest, were:
\begin{itemize}
\item Feedback and PAHs
\item PAH size and charge diagnostics in AGN
\item PAH suppression and destruction radii
\item Cold dust, ices, and aliphatics in obscured AGN
\item Silicate absorption and dust geometry in AGN
\end{itemize}

\section{Summary of the PAH-AGN discussion}
The discussion focused on the role of PAHs and dust as tracers of AGN activity in the JWST era, with particular emphasis on feedback processes and diagnostic power. Participants emphasized that both radiative and mechanical feedback likely contribute to the observed modification of PAH emission in AGN environments, but their relative importance remains uncertain. Future progress will benefit from targeted observations designed to disentangle these effects, for example by combining PAH measurements with ionization and kinematic diagnostics.

JWST observations have revealed a suppression of PAH emission in the vicinity of AGN on $<$~kpc scales; however, whether this reflects true molecular destruction or dilution by an enhanced hot-dust continuum remains an open question. This uncertainty has important implications for the use of PAHs as diagnostics of AGN versus starburst dominance. The intrinsically weak PAH signals in AGN environments also make measurements challenging, and interpretations are often limited by inconsistent fitting approaches and differing theoretical assumptions across studies. As a result, PAHs are likely to be most informative when analyzed within a unified observational and modeling framework and used in conjunction with complementary diagnostics, such as optical emission-line classifications (e.g., BPT diagrams).

\printbibliography[heading=subbibliography]
\end{refsection}

%% file: Structure/Session13.tex
\begin{refsection}
\chapter{Dust grain modeling}

\textbf{Moderator: Frédéric Galliano}\\


\noindent
The term \textit{dust model} can have a relatively wide acceptation. Here we restrain ourselves to the knowledge of the microphysical properties of a grain mixture, that is, for each species:
\begin{itemize}
  \item stoichiometry;
  \item grain structure;
  \item optical properties;
  \item size distribution;
  \item shape distribution;
  \item and abundance relative to the gas phase. 
\end{itemize}
Such a dust model can be used as a framework to interpret observations from the thermal \textit{InfraRed} (IR) emission of a dusty astrophysical object, from its \textit{UltraViolet}-(UV)-to-mid-IR extinction, or from its polarization, \textit{etc.}
It can also be incorporated in numerical simulations of the \textit{InterStellar Medium} (ISM), from protostellar disks to galaxy evolution, determining the extinction, the photoelectric heating rate, or the H$_2$ formation rate, \textit{etc.}

The discussion that we had during the PanDust conference was focused on the main limitations of contemporary dust models, how to use them properly, and formulating the open questions that should pave the way forward.
Although we originally planned to include circumstellar and protoplanetary dust models in the discussion, we stayed mostly focused on ISM dust models.


\section{What do we need to build a dust model?}

\subsection{What are the consensual assumptions?}

Dust models rely on a set of assumptions about the constitution of the grains. 
Contemporary dust models \citep[\textit{e.g.,}][]{Hensley2023,Siebenmorgen2023,Ysard2024} present some similarities and some divergences.
We started our discussion by listing the following similarities.
\begin{itemize}
  \item They are solely constrained by observations from the diffuse ISM of the Milky Way.
  \item The grains are made from a mixture of compounds with a predominance in mass by 
    silicates and carbonaceous materials. 
    Other chemical species (\textit{e.g.,}\ sulfur- or iron-based) can also be included, but
    as minor constituents.
  \item The size distribution must span at least three orders of magnitude, from the 
    size of large molecules (a few Angströms), up to the submicronic range.
\end{itemize}
We then listed the following discrepancies.
\begin{itemize}
  \item The exact composition differs between models, especially concerning the 
    carbonaceous species: PAHs \citep{Hensley2023} or a-C(:H) \citep{Ysard2024}.
  \item The structure is also variable: core/mantle grains with inclusions \citep{Ysard2024}
    or mashups \citep{Hensley2023}. 
    There is also the question of the presence of nanosilicates \citep{Siebenmorgen2023}.
  \item There could be a non-negligible population of micrometric-size grains 
    \citep{Siebenmorgen2023}.
\end{itemize}
A more complete comparison between these different models was reviewed by Ilse \textsc{De Looze}, during the conference.

\subsection{On the exhaustiveness of laboratory studies}

Another important question we discussed is whether enough species have been studied in the laboratory.
More precisely, have the optical properties of the main potential dust analogues been experimentally measured?
We are not sure anyone has the answer to this question. 
However, it seems that:
\begin{itemize}
  \item most of the chemical species found in meteorites \citep[silicates, carbonaceous, 
    oxides, sulfites, silicon carbide, \textit{etc.}; \textit{e.g.,}][]{Hoppe2010} have been 
    studied in the lab. 
  \item most of the commonly observed mid-IR features are attributed to one of these 
    categories, although there are still some debates.
\end{itemize}
One of the issues with the last point is that mid-IR emitting grains are not those dominating the far-IR/submm regime.
The former account for only a small fraction of the total dust mass.
If grains of different sizes indeed have different compositions, then we may be spectroscopically blind to the latter, apart from the few sightlines where we can observe absorption features.

Not all possible silicate stoichiometries nor carbon structures have been studied at all relevant wavelengths, yet.
However, we have the impression that there is already a rich database of laboratory optical properties of solids to design precise dust models,
especially if we account for the virtually limitless ways to combine these solids into composite grains of different structures, sizes and shapes.
It even seems that we currently have too many grain properties compared to the available observables, such as demonstrated by \citet{Zubko2004} and \citet{Ysard2024} who managed to fit the same constraints with different grain mixtures.

\subsection{What constraints should ideally be accounted for, when building a dust model?}
\label{sec:S13:constraints}

All dust modelers agree that, to properly constrain the composition, size and shape distributions of a dust mixture, one needs to account for several of the following observables:
\begin{itemize}
  \item the mid-IR-to-mm continuum emission and polarization;
  \item the far-UV-to-mid-IR extinction curve and polarization;
  \item the elemental depletions;
  \item the albedo of reflection nebulae and diffuse Galactic light;
  \item X-ray absorption edges;
  \item X-ray diffraction halos.
\end{itemize}
Not all these constraints are currently included in contemporary dust models, but they will likely be in the future.

A key point that was discussed is that these constraints need to be coming from the same astrophysical object, without significant confusion.
This is one of the main challenges that will be developed in Sect.~\ref{sec:S13:prospectives}.
This is why the diffuse ISM of the Milky Way has been the target of choice until now.
This is also why it is difficult to design models for external galaxies.
A few attempts were made for the Magellanic clouds, but with a limited set of constraints \citep[\textit{e.g.}][]{Zubko1999,Li2002}.


\section{How to properly use a dust model?}

\subsection{What are the systematic uncertainties of dust models?}

Using a dust model to interpret a set of observations, or including this model into a simulation, both require to assess the different sources of uncertainties it contains.
We could identify the followings:
\begin{itemize}
  \item the uncertainty on the laboratory data, mainly the optical properties, and
    marginally the heat capacity and material density;
  \item the uncertainty on the observations used to constrain the grain properties 
    (emission, extinction, depletions, \textit{etc.});
  \item the uncertainty on the grain parameters derived from fitting the lab data to the
    observational constraints, and the fact that some of these constraints might be 
    poorly fitted;
  \item something more difficult to estimate: the uncertainty on the assumptions made, 
    that is the somehow arbitrary choice of solids, the parametrization of the size and 
    shape distributions, the reference abundance patterns, \textit{etc.}
\end{itemize}
Concerning the last item, we noted that having a suite of dust models, rather than a single one, is not necessarily a bad thing, from an observational point of view.
The dispersion of the results provides a rather natural way to estimate some of the systematic uncertainties on the model assumptions, although it does not provide a complete exploration of all the combinations of parameters.

From a methodological point of view, we noticed that these uncertainties are usually not provided with the dust model by their developers. 
This is understandable, as these are challenging to derive.
An additional difficulty comes from the fact that these uncertainties are often strongly correlated.
For instance, the uncertainties on the continuum emission at two different wavelengths are non-trivially correlated.
A solution to this problem would be to design probabilistic dust models that include this uncertainty at their core. 
In practice, we could imagine providing the user a \textit{Markov Chain Monte Carlo} of all the parameters.
This is what cosmologists do when they demonstrate how a given experiment allows them to refine the estimate of $\Omega_b$, $H_0$, $\sigma_8$, \textit{etc.}

\subsection{How to properly use dust models to interpret observations?}

When using a dust model to interpret observations, the first question is: which model to use? 
In particular, is it better to have a poor fit with a realistic dust model (\textit{e.g.,}\ based on laboratory data) or a good fit with a phenomenological model (\textit{e.g.,}\ a modified black body)?
There was a dissensus about that among our group.
It seems that the answer depends a lot on the scientific goal.
For instance, fitting an IR \textit{Spectral Energy Distribution} (SED) with a modified black body is a way of synthesizing the information contained in the observations (typically a few broadband fluxes) in the form of a few physically-motivated parameters (mass, temperature).
On the other hand, using a realistic dust model allows us to do the same, in addition to assessing how likely the dust model is.
In that sense, having a less flexible, but more realistic model allows us to better see where some of the common hypotheses would break and thus better identify new open questions, while phenomenological models can lead to non-physical fits.
In epistemological terms, we would say that realistic models are more refutable, and thus superior.

The other important question concerns the fitting method. 
There are numerous methods in the literature (frequentist, Bayesian, machine-learning, \textit{etc.}).
They all have their pros and cons.
It is, however, crucial to identify degeneracies between parameters, which are numerous when using dust models.
The Bayesian framework offers a natural way to identify these degeneracies.

\subsection{What do we need in order to go beyond the diffuse ISM of the MW?}
\label{sec:S13:beyondMW}

Since dust properties evolve, from the diffuse to the dense ISM, and also from one galaxy to another, the grain mixture of the diffuse ISM of the Milky Way is an instantaneous snapshot of the particular phase of a particular galaxy at a given time of its evolution.
Using contemporary dust models to interpret external galaxies or denser regions thus likely biases our results.
Ideally, we would need to develop well-constrained models for other environments. 
This would be a way to: \textit{(i)}~understand how the dust parameters (size distribution, composition, \textit{etc.}) vary with environment and \textit{(ii)}~having reliable models to explore other phases or galaxies.

Concerning \textit{local} dust evolution, that is the evolution within the ISM under the effects of gas and radiation field densities, the observational constraints are significantly more complex.
Some observables are challenging to get (\textit{e.g.,}\ the UV extinction toward a dense cloud) and they are more difficult to model, as they usually suffer from confusion.
The diffuse ISM can indeed be reasonably assumed optically thin and uniformly illuminated.
This is not the case for a dense cloud, where detailed radiative transfer is required to make sense of the observations.
It therefore adds another layer of uncertainty.
Colleagues studying dense regions usually extrapolate the grain properties from the diffuse ISM \citep[\textit{e.g.,}][]{Schirmer2020}.

On the other hand, extragalactic dust models are currently at a very preliminary stage. 
The issue is that it is difficult to measure extinction curves outside the Milky Way, and impossible outside the local group. 
Similarly, because of the loss of linear resolution, observations of external galaxies are prone to confusion.
The Magellanic clouds have been used in the past, and are still the best candidates to build fully-constrained extragalactic dust models that would account for the effect of metallicity.


\section{Prospectives}

\subsection{How can computational chemistry advance our knowledge of dust 
   properties and their evolution?}

One of the challenges of dust physics is that we can not yet model grain evolution \textit{ab initio}, combining atoms into molecules and solids, from first principles.
In addition, the computation of precise optical properties is possible only for rather small molecules. 
If such computations were possible for a large collection of species, they would likely allow us to identify the hundreds of \textit{Diffuse Interstellar Bands} \citep[DIBs; \textit{e.g.,}][]{Lallement2024}, which would bring a lot of information on the nature of the small grains.
In addition, we would be able to model grain growth, from small to large grains.
This means we could restrain the wide diversity of possible grain compositions, from the knowledge of the different chemical routes, and not only from spectroscopic information.

Classical computers will not offer this possibility, even in the near-future.
However, quantum computers should allow us to perform such calculations \citep[\textit{e.g.,}][]{Weidman2024}.
Quantum computing is still in its infancy, but it will likely revolutionize our field, and
young researchers interested in astrochemistry should closely follow its development.

\subsection{What observational constraints are needed to progress?}
\label{sec:S13:prospectives}

We reviewed the necessary constraints to build dust models in Sect.~\ref{sec:S13:constraints}, and what is needed to go beyond the diffuse Galactic ISM in Sect.~\ref{sec:S13:beyondMW}.
To get the necessary constraints to build the next generation of dust models that will be applicable to denser regions and external galaxies, we first need the ability to measure reliable UV-visible extinction curves in external galaxies.
This will be possible with HWO.
This observatory will also provide the ability to measure depletions in the same objects.
Second, we will need deep far-IR spectroscopic observations of a diversity of objects, as this spectral range has been poorly explored.
It contains numerous features providing unique information about the crystallinity of silicates or their iron fractions, as well as other compounds.
We might also discover new species.
Besides, compared to mid-IR features, these far-IR features will be mostly carried by the large grains, and will thus uniquely inform us about the composition of the bulk of the mass.
This is why PRIMA \citep[\textit{e.g.,}][]{Moullet2023} will be instrumental for our field.


\printbibliography[heading=subbibliography]
\end{refsection}

%% file: Structure/Session14.tex
\begin{refsection}
\chapter{Dust extinction/attenuation curves in galaxies}

\textbf{Moderator: Samir Salim}\\
\textbf{Additional author: Irene Shivaei}

\section{Extinction curves}

When it comes to the Milky Way dust extinction curves, the principal way of characterizing them is through $R_V$, which represents the slope of the absolute extinction curve in the optical region spanned by $B$ and $V$ filters ($R_V = (A_B/A_V-1)^{-1}$). In that sense, recent development of large-scale $R_V$ maps (e.g., \citealt{zhang24}) represents a big step forward. These maps allow us to identify the trends in $R_V$ distribution and make connections to other properties, such as the PAH distribution \citep{lee25}, which should help us get a fuller understanding of the grain size evolution. However, there are open questions regarding $R_V$ maps. One is understanding the resolution effects, i.e., how much variation there is at different spatial scales and, ultimately, in the limit of point sources, how good is the agreement with $R_V$ values from the individual sightlines.

$R_V$ maps probe typical diffuse ISM regions, where the variation of $R_V$ is relatively small. The critical test of dust properties comes from sightlines with more extreme $R_V$ values, or otherwise less usual curves (e.g., weak UV bumps). Currently, the number of such sightlines in the diffuse ISM is limited. The hope is that this might change in the future, but this might not happen before some new UV facilities become available (e.g., the proposed UVESS satellite described in a poster by Battisti et al.).

Recently some important advances have been made to increase the number of extinction sightlines beyond the MW \citep{gordon24}. The principal goal of such studies is to identify  the second parameter that controls the shape of extinction curves beyond $R_V$. This might be the dust-to-metals ratio. However, the measurements of metals in gas-phase are difficult. Overall, challenges remain in getting these datasets to be as rich as they are for the MW. Even larger challenges lie in extending the work to beyond the Local Group, e.g., to include galaxies with very low metallicities. Ultimately, these goals should be achieved with the Habitable Worlds Observatory (HWO, see \citealt{dressing26} for a science case about the extinction curves), but perhaps some significant advances might come earlier from UVESS or UVEX. 

\section{Extinction vs.\ attenuation curves}

Extinction curves represent the starting point for understanding the attenuation curves, which can be drastically different from extinction curves due to the radiative transfer (RT) effects and the differences in the distribution of stars and dust (the ``geometry''). A potential way forward would be to expand the number of galaxies and sightlines where one could study both the extinction and the attenuation. By studying the differences between the two we might learn about the geometry and RT effects. One possibility would be to use the extinction curves of the optical counterparts of gamma ray bursts (GRBs) and constrain the attenuation curves of their hosts. This, however, might be challenging since most GRB hosts are quite faint. Also, reliable individual attenuation curves require mid or far-IR photometry, which are not readily available, but should be with PRIMA.

Another open question related to the relationship between extinction and attenuation curves regards the spatial scale at which the geometry and RT effects start to significantly modify extinction into attenuation. This is something that could potentially be addressed using models.

On the very practical level, the discussion also touched upon the confusion arising from using the same nomenclature for extinction and attenuation (e.g., $E(B-V)$), but the participants did not have the time to come up with any proposed solutions.

\section{Attenuation curves}

We now know that the attenuation curves, and especially their slopes, vary significantly from one galaxy to another. What is less well known is how much of this variation is due to the differences in the local arrangement between dust and stars in galaxies as opposed to, for example, the viewing angle. Again, we might need to utilize realistic simulations to address this question, instead of using just the toy models.

Related to this is the question of whether there are systematic differences in the  geometry between different galaxy types (e.g, mergers). This would be hard to determine because other factors, including the intrinsic dust properties, affect the attenuation curve as well, coupled with the fact that the constraints on the shape (or even just the slope) of attenuation curves are not very precise.

The previous point is also related to the methodologies of measuring dust attenuation curves. We should do more to understand the caveats and potential biases of both the empirical and the SED fitting methods. In regards to the empirical method, it is the fact that it produces aggregate curves that might be a source of limitations. As for the SED fitting, in addition to the assumptions regarding the star formation histories, it might rely on the application of the energy balance, which is not usually tested. Again, the simulations may help.

\printbibliography[heading=subbibliography]

\end{refsection}

%% file: Structure/Session15.tex
\begin{refsection}
\chapter{High redshift dust}

\textbf{Moderator: Jed McKinney}\\
\textbf{Additional author: Emma Lieb}

\section{An interactive discussion on dust in the early Universe}
\begin{figure}[h]
    \centering
    {{\includegraphics[width=0.45\textwidth]{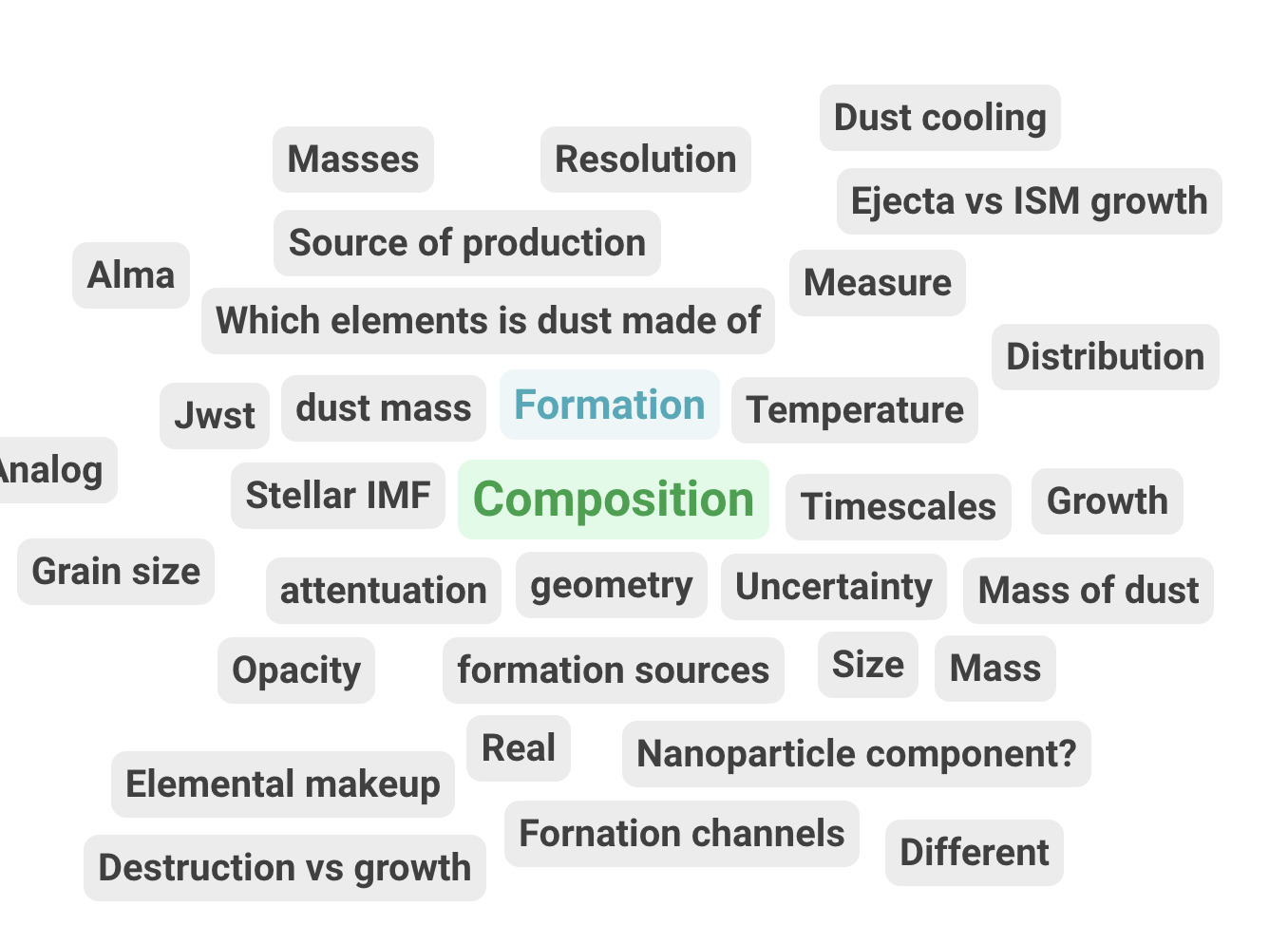} }}%
    {{\includegraphics[width=0.54\textwidth]{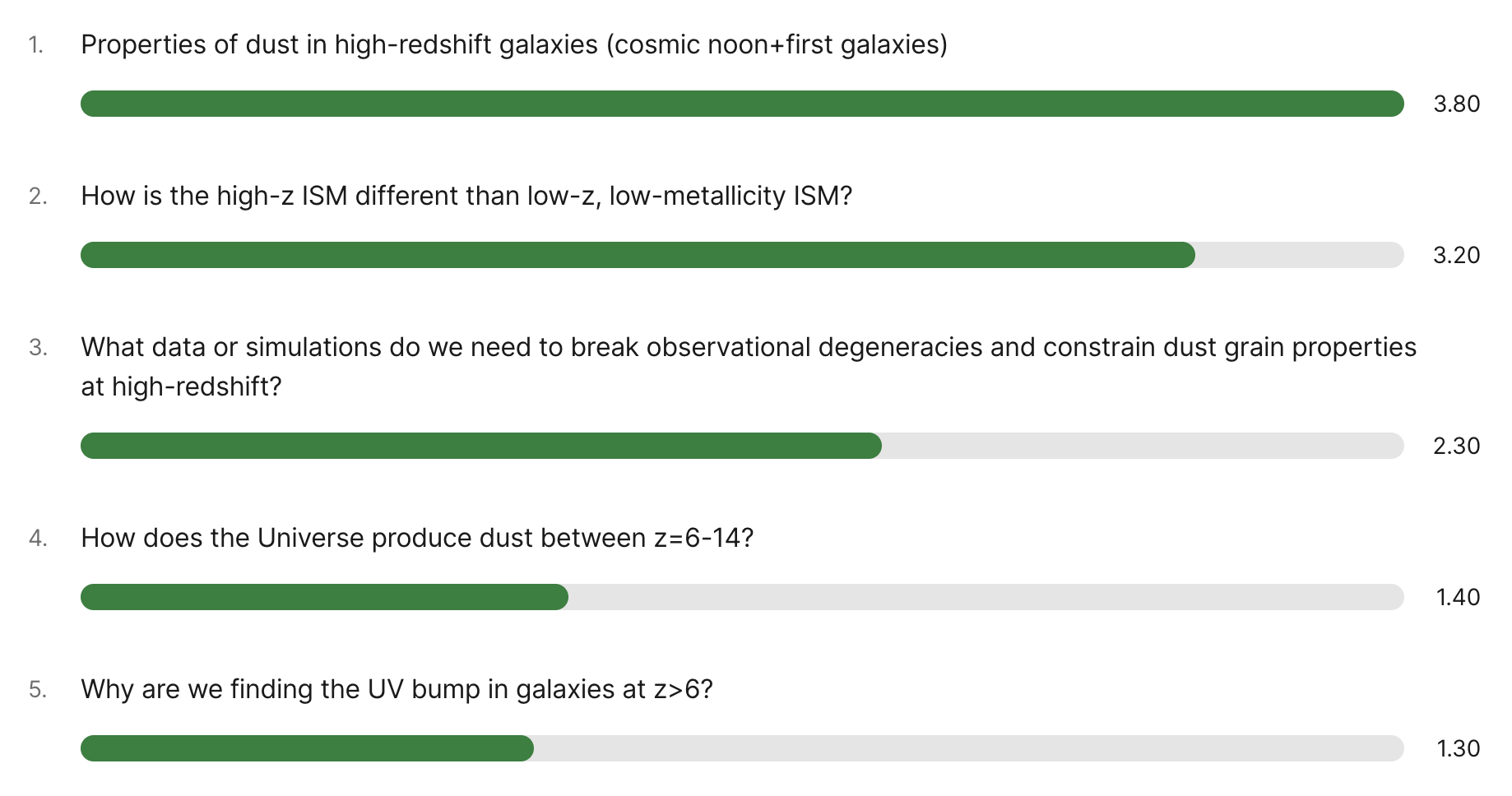} }}%
    \caption[]{Interactive discussion results using \href{https://www.slido.com/}{slido} to collect audience participation on-the-fly. (\textit{Left}) Word cloud based on responses to ``What comes to mind when thinking of high-redshift dust?''. (\textit{Right}) Top 5 community-driven questions proposed and voted for. 
    }%
    \label{fig:section15:interactiveSection}%
\end{figure}

The amount of dust and its properties remain a confounding issue in the broader study of high-redshift galaxies. For the purposes of this discussion, we defined high-redshift to be $z>6$. Using \textit{slido}, an audience interaction tool, this discussion group began by brainstorming key terms that came to mind when thinking of high-redshift  dust. As shown on Figure \ref{fig:section15:interactiveSection} (\textit{Left}) these touch on the properties of dust grains (e.g., ``composition'') as well as the mechanisms that produce reservoirs of dust (e.g., ``formation, timescales''). Next, the group voted on more specific topics to discuss, the top 5 of which are shown in Figure \ref{fig:section15:interactiveSection} (\textit{Right}). Using these as a guide, the remaining discussion can be broadly summarized around (1) the properties of high-redshift dust, and (2) the sources of high-redshift dust. 

\section{Summary of discussion on the properties of high-redshift dust}
From the perspective of a few participants, ``high-redshift dust should be the same as local dust'' as it is made of the same elements. In some regards, this is the best case scenario as it provides a pathway for obtaining the most meaningful constraints on high redshift dust and properly accounting for its effect on observed SEDs and in its regulation of star-formation. However, the in-situ environments of high redshift galaxies exhibit key differences from local galaxies, including mixing, the burstiness of star formation, and radiation fields to name a few that came up in the discussion. The group acknowledged studying intermediate redshift ($z\sim1-2$) galaxies as a stepping stone, but these more evolved systems tend to exhibit complex stellar and dust geometries with complex impact on the effects of dust attenuation. This might be somewhat less problematic at higher redshifts where galaxies and their star formation are more compact. 

On the other hand, there are some reasons to assume that the dust in $z>6$ galaxies has some divergent properties relative to local dust. A likely driver of such differences would be dust formation from supernovae as opposed to AGB stars that tend to dominate dust production at lower redshifts. The grain size distributions and compositions from supernovae both before and after reverse shock processing remain a significant barrier to interpreting such dust, and are therefore a promising area for future research. Indirect evidence favors shallow rest-frame UV attenuation laws at high redshift, pointing towards grain size distributions weighted towards larger particles. Recent simulations suggest that the observables of dust at $z>10$ are largely shaped by supernovae dust production, and $z=10\rightarrow6$ marks the transition to dust being shaped by grain growth mechanisms at play in the ISM. 


\section{Summary of discussion on the sources of high-redshift dust}
Our discussion began with investigating what are largely considered the three main sources of cosmic dust (at all redshifts): Asymptotic Giant Branch (AGB) stars, Supernovae (SNe), and the growth of dust grains in the ISM. While AGB stars may produce the most carbonaceous dust in the local universe, it is unlikely for them to be major sources at high-redshift due to their low initial masses and therefore long timescales for dust production. Galaxies at high-redshift require their stellar dust sources to be of high-mass which are able to produce dust on short timescales. It is also true that there are more high-mass stars as a function of redshift. This makes it more likely for SNe to be a major source of dust at high redshifts; SNe have the high initial masses required for dust production in early galaxies. However, these SNe primarily create silicate dust rather than carbonaceous dust. This brings us back to the question of whether or not local dust is a good laboratory for high redshift dust. If we can say that the dust at high redshift should be the same dust in our local environment, i.e. if all dust is created under the same rules of physics, then we would expect the mixing of dust at high-redshift to be comparable to the dust in our local universe. This would mean a population of both silicate and carbonaceous materials (both either crystalline or amorphous) as well as Polycyclic Aromatic Hydrocarbons (PAHs). ISM growth is therefore a strong contender for the prominent source of dust at high-z. ISM growth requires no stellar lottery to be won in terms of the necessary variables to be at play (initial mass functions, metallically, etc.). One minor talking point brought up an often-overlooked source of carbonaceous dust at high-z which is Wolf-Rayet (WR) binaries, which have been shown to create carbonaceous dust in the MW and have the short-lifetimes required to reach their dust formation phase in the early universe. These systems have also been shown to create PAHs. While we have no direct evidence for PAHs at high-redshifts, some observations of TMC 1 have shown highly shielded simple molecules on which PAHs can grow. Our discussion did not land on one source of dust winning out over any others. It seems highly probable that dust in high-z galaxies is seeded by stellar events and then grows in the ISM. This question will require further observations from facilities like JWST to confirm the full lifecycle of dust grains at high-redshift. 





\printbibliography[heading=subbibliography]
\end{refsection}

%% file: Structure/Session16.tex
\begin{refsection}
\chapter{The future of studying dust}

\textbf{Moderator: J.D. Smith}

\section{Introduction}
Historically, projections into the future of any area of research age very poorly.  From “\textit{flying cars by 2020}”, to population bombs, to energy “\textit{too cheap to meter}”, to Lord Kelvin’s confident pronouncement made just prior to the age of relativity and quantum that “\textit{There is nothing new to be discovered in physics now}”, the present very often proves an obscured vantage from which to survey the near and distant future. And so it is with the study of the condensed phase of heavy elements in the Universe — of dust. New tools and techniques, sophisticated models, fundamental theory, and paradigm-shattering observations always lie tantalizingly just around the corner.  So rather than concrete predictions for the future, we framed our discussion around the following thought experiment:

\vspace{1em}
\begin{adjustwidth}{1em}{1em}
You are a graduate student studying dust in the year 2075.  What are the durable results in the field you build your research on?
\end{adjustwidth}

\section{Discussion}

One clear consensus that emerged is that the role of dust in studying the key physical processes in the Universe --- from star and planet formation to galaxy evolution to life itself --- grows year by year, and that this trajectory seems quite durable. And yet, while the detailed study of dust and the important role it plays in many cosmic ecosystems has been pushed forward considerably over the past several decades by groundbreaking facilities including Hubble, ISO, Spitzer, Herschel, ALMA and now JWST, it is also clear that large scale, uniform surveys (SDSS, GAIA, DESI, etc.) which leveraged the statistical power of millions of targets in massive, coherent datasets have produced a qualitatively different type of science.  Such surveys have teleported the study of star formation, Galactic structure, metal enrichment, and stellar mass assembly in the Universe forward in several remarkable lurches.  Measuring the detailed ISM of every galaxy with “\textit{an infrared SDSS}”, probing the composition and 3D structure of dust in systematic volume-complete samples of Galactic sources and lines of sight with “\textit{an IR spectroscopic GAIA}”, obtaining an extinction curve for every O star in the Galaxy --- all expressions of a yearning within our field for similar quantum jumps.

The driving role of instrumentation in determining what experiments we can bring to bear on the study of dust is also key.  Those who pursue dust in  Milky Way sight-lines must continuously plead the case for high dynamic range surface brightness sensitivity in new platforms and instruments.  Important physical insights often require key capabilities like infrared polarimetry which are, for the most part, simply missing from the full repertoire of astronomical instrumentation. Even ALMA is challenged in the study of dust at very high redshift. Large scale surveys of environments from the Milky Way to the early Universe with future facilities like Roman (if successful) and PRIMA (if selected) will be pivotal, with the potential to move dust astrophysics further into the grand survey regime.

Equally important are the tools, techniques, and surrounding capabilities we cultivate in our field.  Simulations of everything from proto-planetary disks to evolving populations of galaxies are rapidly improving the fidelity of the underlying chemical pathways, dust life cycles, and radiative transfer, reaching the point that they can be used as experiments for evaluating the broad brushstrokes of dust microphysics (see Session 6).  A continuous cycle of \textit{productive friction} between simulations and observations is the vital next step.  At another level in the physical scale abstraction, carefully exploring bottom-up reactions — building dust grains using realistic theoretical and numerical experiments from individual atoms and parent grains in differing contexts — will provide key “reality testing” of the sub-grid physics used in larger scale astrophysical dust simulations.  The rapid development and application in astronomy of tools of Machine Learning will likely prove pivotal in facilitating these connections, though interpretability and reliability must be baked in from the beginning.

But beneath the observational tools and techniques, the theory and its foundations, lies perhaps the most important aspect of our field: the \textit{culture} of studying dust.  While many sub-fields of astronomy have thriving research cultures, those who study dust are a special breed, united by the basic need to build community and answer the question “\textit{Why dust?}”.  Making the most of what we have on hand, with strong support for archival work, improved opportunities for collaboration, and a spirit of sharing the skills, datasets, models, and frameworks which lie at the ragged edge of our field were seen as vital.  Younger practitioners are keen to bridge the gaps between generations and wavelength regimes, emphasize better hiring standards, and work to increase opportunities for collaboration.  Reduced proprietary time, improved access to student training in key areas of data science, central repositories of relevant lab data into which many groups are incentivized to feed and organize their most useful results — all example areas where greater emphasis would pay strong dividends.  

When asked ``\textit{If you had your life to live over, would you be an astronomer again?}'' the astronomer Walter Baade famously replied: ``\textit{Only if the ratio of total to selective absorption is everywhere the same.''} \citep{1968Obs....88..168V}  In the more than 50 years since the question was asked, dust has moved from nuisance, to curiosity, to niche topic, to the vital component of our understanding of the Universe it is today.  While we cannot know what the next 50 years will bring, one thing is clear: dust will continue to reveal fundamental insights about our cosmos.

\printbibliography[heading=subbibliography]
\end{refsection}